\documentclass[aps,prd,showpacs,twocolumn,
superscriptaddress]{revtex4}

\usepackage{graphicx}
\usepackage{dcolumn}
\usepackage{doi}
\usepackage{amsmath}
\usepackage{amssymb}
\usepackage{bm}
\usepackage{color}
\usepackage[normalem]{ulem}
\usepackage[dvipsnames]{xcolor}
\usepackage{hyperref}
\hypersetup{colorlinks=true,linkcolor=blue,
citecolor=cyan,}

\begin{document}
\title{Test particle motion around a black hole in Einstein-Maxwell-scalar theory}
\author{Bobur Turimov}
\email{bturimov@astrin.uz}
\affiliation{Ulugh Beg Astronomical Institute, Astronomicheskaya 33, Tashkent 100052, Uzbekistan }\affiliation{Institute of Physics, Faculty of Philosophy and Science, Silesian University in Opava, Bezru\v covo n\' am. 13, CZ-74601 Opava, Czech Republic}
\author{Javlon Rayimbaev}
\email{javlon@astrin.uz}
\affiliation{Ulugh Beg Astronomical Institute, Astronomicheskaya 33, Tashkent 100052, Uzbekistan }
\affiliation{National University of Uzbekistan, Tashkent 100174, Uzbekistan}
\author{Ahmadjon~Abdujabbarov}
\email{ahmadjon@astrin.uz}
\affiliation{Ulugh Beg Astronomical Institute, Astronomicheskaya 33, Tashkent 100052, Uzbekistan }
\affiliation{Shanghai Astronomical Observatory, 80 Nandan Road, Shanghai 200030, People's Republic of China}
\affiliation{Tashkent Institute of Irrigation and Agricultural Mechanization Engineers, Kori Niyoziy, 39, Tashkent 100000, Uzbekistan}
\affiliation{National University of Uzbekistan, Tashkent 100174, Uzbekistan}

\author{Bobomurat Ahmedov}
\email{ahmedov@astrin.uz}
\affiliation{Ulugh Beg Astronomical Institute, Astronomicheskaya 33, Tashkent 100052, Uzbekistan }

\affiliation{Tashkent Institute of Irrigation and Agricultural Mechanization Engineers, Kori Niyoziy, 39, Tashkent 100000, Uzbekistan}
\affiliation{National University of Uzbekistan, Tashkent 100174, Uzbekistan}

\author{Zden\v ek Stuchl\' ik}
\email{zdenek.stuchlik@fpf.slu.cz}
\affiliation{Institute of Physics, Faculty of Philosophy and Science, Silesian University in Opava, Bezru\v covo n\' am. 13, CZ-74601 Opava, Czech Republic}

\date{\today}
\begin{abstract}

In this paper, we explore the test particle motion around black hole in Einstein-Maxwell-scalar (EMS) theory using three different black hole solutions within this theory. We have first analyzed the spacetime curvature structure of these solutions and shown the existence of two singularities and the first one is at the center $r=0$. In black hole spacetime, there are two regions divided by the critical value of the cosmological parameter $\lambda_0$. The photon sphere around the black hole in EMS theory has also been studied and found that it does not depend on cosmological parameter $\lambda$. We have analyzed the innermost stable circular orbits (ISCO) around the black hole and shown that for all solutions ISCO radius for neutral particle decreases with the increase of black hole charge. We have also studied the charged particle motion around the black hole where charged particle motion is considered in the presence of the gravitational field and the Coulomb potential. It is shown that ISCO radius for charged particles increases depending on the selected value of the coupling parameter which is in contradiction with observations of the inner edge of the accretion disks of the astrophysical black holes and can be used as a powerful tool to rule out the EMS theory from consideration for the gravitational field theory. It is also studied the fundamental frequencies governed by test particle orbiting around the black hole in EMS theory. Finally, as a test of black hole solution in EMS theory ISCO radii is compared with that in Kerr black hole and found that the spin parameter of Kerr can be mimic up to $a/M\simeq 0.936$. 

\end{abstract}

\maketitle

\section{Introduction}

%%%%%%%%%%%%%%%%%%%%%

In the low energy limit of string theory, one may introduce the dilaton scalar field which is appearing as an additional supplement term to the Einstein action in the form of the axion, gauge fields, and another nontrivial coupling of dilaton to fields. 
The causal structures and thermodynamic properties of black hole solution with dilaton have been explored in~\cite{Gibbons88,Garfinkle91,Brill91,Gregory93a,Koikawa87,Boulware86,Rakhmanov94,Harms92, Holzhey92}.

Another interesting feature comes from study of the black hole solutions with the cosmological constant. The correspondence of anti-de Sitter solution with conformal field theory may be used to unify quantum fields and gravitons. Theories containing negative cosmological constants may be considered as a part of supergravity theories defined in higher-dimensional spacetime. The study of these type theories and solutions can be found in Refs.~\cite{Maldacena98, Maldacena99, Witten98, Klemm01, Gubser98, Aharony00}.

In heterotic string theory, the scalar dilaton field is coupled to higher-order terms of module of electromagnetic field tensor. Consequently, for nonzero electromagnetic field tensor, one may have nonconstant dilaton term and in the Reissner-Nordstr\"{o}m limit the solution is not the approximate solution of the string theory~\cite{Garfinkle91}. In Refs.~\cite{Jai-akson2017,Heydari-Fard2020}, the black hole merger estimation and thin accretion model within framework of Einstein- Maxwell-dilaton gravity has been investigated.

Existing black hole solutions can be tested using test particle motion around gravitational compact objects in the strong gravity regime. Standard solutions such as Schwarzschild and Kerr black hole ones within general relativity have been successfully tested in both weak (using solar system tests)~\cite{Eubanks97,Iorio10,Lobo10,Olmo11} and strong field regime (observation of gravitational waves, the shadow of black holes, and stars moving around the supermassive black hole Sagittarius A* (SgrA*) at the center of Milky Way galaxy)~\cite{LIGO16,EHT19a,EHT19b} and GRAVITY instrument. However, in these observations, there is still open window for testing other theories of gravity, including the black hole solution with the dilaton scalar field with the cosmological constant. Particularly, the x-ray observation data from astrophysical compact objects have been used to test alternative and modified theories of gravity in Refs.~\cite{Bambi16c,Zhou18,Tripathi19}. 

It is widely believed that the quasiperiodic oscillations (QPOs) observed around gravitational compact objects is one of the promising tools to test the phenomena occurring in the strong gravitational field of the black hole candidates observed as x-ray microquasars. One of the developed models to explain such observational phenomena is the epicyclic frequencies governed by neutral test particle motion orbiting around the black hole, for example, ~\cite{Abramowicz2001,Torok2005,Torok2005b,Stuchlik2013}, by the charged particle motion in Refs.~\cite{Tursunov16,Kolos17}.

The circular orbits of test particles, particularly innermost stable circular orbits (ISCOs) are the subject of special interest. The observation of accretion disc may be used to get estimation and constraints on parameters of black hole~\cite{Steiner11,Gou14,McClintock14,Steiner10}. Particularly, the presence of the magnetic field around black holes changes the structure of the dynamics of charged particles~ \cite{Chen16,Hashimoto17,Dalui19,Han08,Moura00,Morozova14,Stuchlik2020Univ}. The studies of spacetime structure and particle motion around black holes may be found in Refs.~\cite{Pugliese10,Pugliese11,Jawad16,Hussain15,Jamil15,Hussain17,Babar16,Banados09,Majeed17,Zakria15,Brevik19,DeLaurentis2018PhRvD,Turimov18a}.  

Another possible way of study of the particle dynamics is connected with magnetized particle motion around the black hole in the external magnetic field ~\cite{deFelice,defelice2004}. 
Recently the magnetized particle motion around the non-Schwarzschild black hole has been studied in~\cite{Rayimbaev16}. A similar study in the presence of the quintessence parameter has been carried out in ~\cite{Oteev16}. Magnetized particle motion as a test of spacetime structure and gravity theories have been explored in Refs.~\cite{Toshmatov15d,Abdujabbarov14,Rahimov11a,Rahimov11,Haydarov20,Haydarov20b,Rayimbaev20204DEGB}.
The electromagnetic field structure and charged particle dynamics around black hole have been widely studied in the literature~\cite{Kovar10,Kovar14,Aliev89,Aliev02,Aliev86,Frolov11,Frolov12,Stuchlik14a,Shaymatov14,Abdujabbarov10,Abdujabbarov11a,Abdujabbarov11,Abdujabbarov08,Karas12a,Shaymatov15,Stuchlik16,Rayimbaev20,Turimov18IJMPD,Turimov18b,Shaymatov18,Turimov17,Shaymatov20egb,Rayimbaev15,shaymatov19b,Rayimbaev19,Shaymatov20b,Narzilloev2020C,Narzilloev19,Rayimbaev2020PRD}.

In this paper, we are interested in studying test particle motion in the spacetime of the Einstein-Maxwell-scalar theory. The paper is organized in the following way. In Sec. ~\ref{Solution}, we provide in very detailed form the exact analytical solutions of the Einstein-Maxwell-scalar (EMS) field equations. Section ~\ref{particlemotion} is devoted to the derivation of test particle motion. Finally, in Sec. ~\ref{Summary}, we summarize obtained results and give future outlook related to the present work.

Throughout the paper, we use a spacelike signature $(-,+,+,+)$, a system of units in which $G = c = h= 1$ and restore the Newtonian constant, speed of light and plank constant when we need to compare the results with the observational data. Greek indices are taken to run from $0$ to $3$, and Latin indices from $1$ to $3$.

%%%%%%%%%%%%%%%%%%%%%%%%%%%%%%%%%%%%%%%%%%%%%%%%%%%%%%

\section{Spacetime metric\label{Solution}}

In this section, we plan to incorporate into the Einstein-Maxwell-scalar field equations. The action for such a system can be described by ~\cite{Gibbons88,Garfinkle91,Yu20}
\begin{align}\nonumber\label{action}
S=\int d^4x\sqrt{-g}\Big[&R-2\nabla_\alpha\phi \nabla^\alpha\phi\\&-K(\phi) F_{\alpha\beta}F^{\alpha\beta}-V(\phi)\Big]\ ,
\end{align}
where $\nabla_\alpha$ stands as covariant derivative, $g$ is the determinant of the metric tensor $g_{\mu\nu}$, $R$ is the Ricci scalar of the curvature and $\phi$ is the massless scalar field, $F_{\alpha\beta}$ is the electromagnetic field tensor, and $K(\phi)$ is the coupling function between the dilaton and the electromagnetic fields. $V(\phi)$ is the scalar potential. 

Varying the action (\ref{action}) with respect to the metric $g_{\alpha\beta}$, vector potential $A_\alpha$, and dilaton (scalar) $\phi$ fields, respectively, we obtain equation of motion of EMS system, namely,~\cite{Yu20} 
\begin{align}\label{eq1}
\nonumber&R_{\alpha\beta}=2\nabla_\alpha\phi\nabla_\beta\phi+\frac{1}{2}g_{\alpha\beta}V\\&+2K\left(F_{\alpha\gamma}F_\beta^{\,\,\gamma}-\frac{1}{4}g_{\alpha\beta}F_{\mu\nu}F^{\mu\nu}\right)\ ,
\\\label{eq2}
&\nabla_\alpha\left(KF^{\alpha\beta}\right)=0\ ,
\\\label{eq3}
&\nabla_\alpha\nabla^\alpha\phi-\frac{1}{4}\left(V_{,\phi}+K_{,\phi}F_{\mu\nu}F^{\mu\nu}\right)=0
\end{align}
where $_{,\phi}$ denotes derivative with respect to the dilaton field, $R_{\alpha\beta}$ is the Ricci tensor. 

One can see that the equations (\ref{eq1})-(\ref{eq3}) are coupled with differential equations and it is difficult to find exact solutions of such a system of coupled nonlinear equations. Nevertheless, it one can find exact analytical solutions for the complex systems under some assumptions. The general form of exterior metric for the spherically symmetric and static black hole can be written as~\cite{Yu20}
\begin{align}\label{metric}
ds^2=-U(r)dt^2+\frac{dr^2}{U(r)}+f(r)\left(d\theta^2+\sin^2\theta d\varphi^2\right)\ ,
\end{align}
where $U(r)$ and $f(r)$ are the unknown radial metric functions. Assume the time component of the vector and the dilaton fields depend on the radial coordinate only, i.e., $A_t=A_t(r)$ and $\phi=\phi(r)$.

In the present paper, we will not look for the new solutions of EMS equations; however, we will test some exact solutions obtained by other authors. In the paper, we will consider three different black hole solutions. The first one is Garry-Maeda-Garfinkle-Horowitz-Strominger (GMGHS) solution~\cite{Gibbons88,Garfinkle91},
\begin{align}
&K=e^{2\phi}\ ,\quad V=0\ ,   
\end{align}
the second one is Gao-Zhang (GZ) solution~\cite{Gao04},
\begin{align}
&K=e^{2\phi}\ ,\quad V=\frac{\lambda}{3}\left(e^{2\phi}+4+e^{-2\phi}\right)\ , 
\end{align}
and finally, the last one is Yu-Qiu-Gao (YQG) solution~\cite{Yu20},
\begin{align}\label{KV}
&K=\frac{e^{2\phi}}{\beta e^{2\phi}+\beta-2\gamma}\ ,\quad V=\frac{\lambda}{3}\left(e^{2\phi}+4+e^{-2\phi}\right)\ .
\end{align}
\begin{table*}
\caption{\label{Tab1} List of radial functions for three different solutions of EMS theory are presented}
\begin{ruledtabular}
\begin{tabular}{ccccc}
Solutions & $f(r)$ & $U(r)$ & $A_t$ & $\phi(r)$ \\
\hline
GMGHS~\cite{Gibbons88,Garfinkle91} & $r^2\left(1-\frac{Q^2}{Mr}\right)$ & $1-\frac{2M}{r}$ & $-\frac{Q}{r}$& $-\frac{1}{2}\ln\left(f/r^2\right)$ \\\\
GZ~\cite{Gao04} & $r^2\left(1-\frac{Q^2}{Mr}\right)$ & $1-\frac{2M}{r}-\frac{\lambda}{3}f$ & $-\frac{Q}{r}$& $-\frac{1}{2}\ln\left(f/r^2\right)$ \\\\
YQG~\cite{Yu20} & $r^2\left(1+\frac{\gamma Q^2}{Mr}\right)$ & $1-\frac{2M}{r}-\frac{\lambda}{3}f+\frac{\beta Q^2}{f}$ & $\frac{Q}{r}\left[\gamma-\frac{\beta}{2}\left(1+\frac{rQ}{f}\right)\right]$& $-\frac{1}{2}\ln\left(f/r^2\right)$ \\
\end{tabular}
\end{ruledtabular}
\end{table*}

The unknown radial functions $f(r)$, $U(r)$, $\phi(r)$, and $A_t(r)$ corresponding to the three different solutions are listed in Table~\ref{Tab1}. One can see that among them YQG solution is more general and characterized by five independent free parameters: black hole mass $M$ and charge $Q$, the cosmological parameter $\lambda$, and additional two $\gamma$ and $\beta$ parameters arisen from modification of $K(\phi)$ function as shown in Eq. (\ref{KV}). It covers the other two solutions, for example, in the case when $\gamma=-1$ and $\beta=0$, YQG solution reduces to GZ solution and then switching off $\lambda$ parameter we obtain the well-known GMGHS solution.   

Using the standard procedure, the dilaton charge $D$, the electric charge $Q$, and the anti-de Sitter mass $M$ of the black hole can be calculated by~\cite{Yu20}
\begin{align}
&D=\frac{1}{4\pi}\int\nabla_\alpha\phi \,_{*}dS^\alpha\ , 
\\
&Q=\frac{1}{4\pi}\int K(\phi)\nabla_\alpha A_t \,_{*}dS^\alpha\ ,
\\
&M=\frac{1}{4\pi}\lim_{S^\alpha\to i^0}\int g^{\mu\nu}\left(\partial_\beta g_{\mu\alpha}-\partial_\mu g_{\alpha\beta}\right)\,_{*}dS^\alpha \ ,
\end{align}
where $_{*}dS^\alpha$ is dual element of the hypersurface $dS^{\alpha,\beta,\gamma}$ , $i^0$ is the spacetlike infinity, and $\partial_\alpha$ stands for partial derivative with respect to coordinate $x^\alpha$.

In order to better understand the structure of the spacetime geometry in Eq.~(\ref{metric}), we determine the curvature invariants such as Ricci scalar $R$, Ricci square $R_{\alpha\beta}R^{\alpha\beta}$, and the Kretschmann ${\cal K}=R_{\alpha\beta\mu\nu}R^{\alpha\beta\mu\nu}$ scalar. Before starting to calculate them, we use the following fact that structure of $f(r)$ function is the same in all three solutions which satisfies to the following relations: 
\begin{align}\label{f}
f'(r)=r+\frac{f(r)}{r}\ , \qquad f''(r)=2\ .    
\end{align}
Taking into account Eq. (\ref{f}) and hereafter performing simple algebraic calculations, the curvature invariants for the general form of spacetime metric (\ref{metric}) can be expressed as 
%
%\begin{widetext}
\begin{align}\nonumber
&R=\frac{Ur^2}{2f^2}+\frac{2-2rU'-3U}{f}+\frac{U}{2r^2}-\frac{2U'}{r}-U''\ ,
\\\nonumber
&R_{\alpha\beta}R^{\alpha\beta}=\frac{U^2r^4}{4f^4}-\frac{r^2 U \left(rU'+2U\right)}{2f^3}+\frac{A}{2f^2}+\frac{B}{2r^2 f}\\\nonumber&+\frac{U^2}{4r^4}-\frac{UU'}{2r^3}+\frac{2U'^2-UU''}{2r^2}+\frac{1}{2}U''^2+\frac{U'U''}{r}\ , 
\\\nonumber\label{K}
&{\cal K}=\frac{3U^2r^4}{4f^4}-\frac{r^2U\left(r U'+U+2\right)}{f^3}\\&+\frac{r^2 U'^2+U\left(rU'-4\right)+\frac{9 U^2}{2}+4}{f^2}\\\nonumber&+\frac{2r^2U'^2+rU U'-U(U+2)}{f}\\\nonumber&+\frac{3U^2}{4r^4}-\frac{UU'}{r^3}+\frac{U'^2}{r^2}+U''^2\ ,    
\end{align}
where 
\begin{align*}
&A=4+7 U^2-U\left(r^2U''-5rU'+8\right)+2 rU'\left(rU'-2\right)   
\\
&B=2rU'\left(r^2 U''+2 rU'-2\right)+rU\left(2rU''+5 U'\right)-2 U^2 .
\end{align*}

Here we do not show the explicit form of the expressions for the curvature invariants for all three black hole solutions; however without losing generality, our analyses show that in each black hole solution all the curvature infuriates in (\ref{K}) have two singularities one is $r=0$ and other one appears at $f=0$ which is corresponds to a naked singularity. 

In the case, when $Q=0$, the curvature invariants in (\ref{K}) take the form 
\begin{align}\nonumber\label{limit}
&\lim_{Q\to 0}R=4\lambda\ ,
\\
&\lim_{Q\to 0}R_{\alpha\beta}R^{\alpha\beta}=4\lambda^2\ ,
\\\nonumber
&\lim_{Q\to 0}{\cal K}=\frac{48M^2}{r^6}+\frac{8\lambda}{3}\ .
\end{align}
From the expression (\ref{limit}), one can see that despite the absence of the black hole charge Ricci scalar and Ricci square are nonzero for YQG solution. An interesting fact is that in the case when $Q=0$, the curvature invariants are independent of $\gamma$ and $\beta$ parameters. Here one should emphasize that the results presented in (\ref{limit}) will also be the same in GZ solution. Finally, in the case when $Q=0$ and $\lambda=0$ the curvature invariants reduce to the Schwarzschild case as $R=R_{\alpha\beta}R^{\alpha\beta}=0$ and ${\cal K}=48M^2/r^6$. 

\begin{figure*}
\includegraphics[width=\hsize]{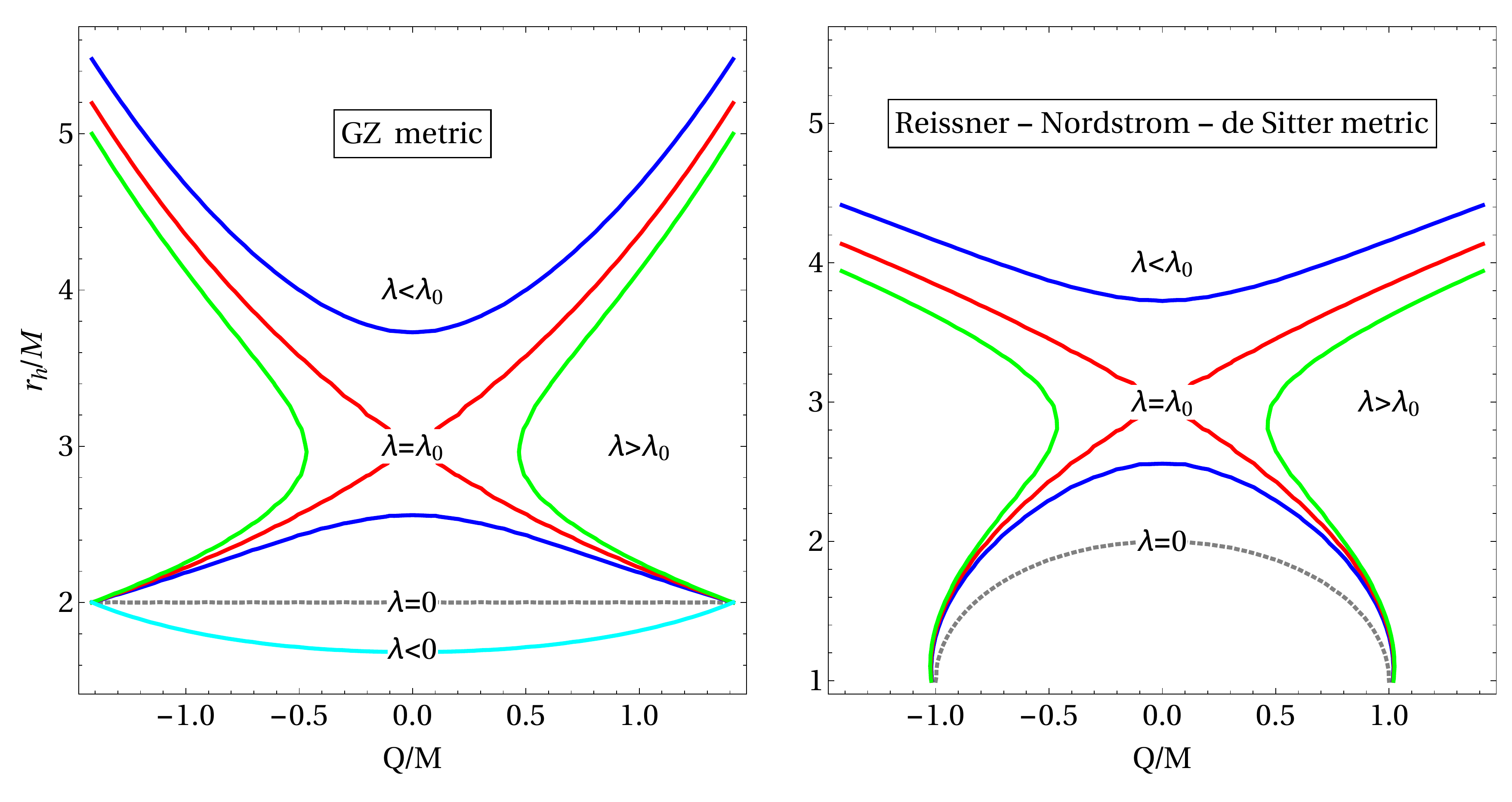}
\caption{(Color online) Radius of the horizon for GZ metric (left) and Reissner-Nordstr{\"o}m-de Sitter metric (right) are functions of black hole charges for the different values of $\lambda$ parameter.\label{fig:horizon}}
\end{figure*}

The radius of the horizon is a quite common phenomenon in spacetime structure and it can be found from $U=0$ for all three black hole solutions. For GMGHS metric, it coincides with the Schwarzschild radius, i.e., $r_h=2M$, while for GZ metric, horizon radius can be found as a solution of the following polynomial equation:
\begin{align}\label{rh}
1-\frac{2M}{r}-\frac{\lambda}{3}r^2\left(1-\frac{Q^2}{Mr}\right)=0\ . 
\end{align}
The introduced spacetime metric is an interesting generalization of the Schwarzschild-de Sitter one governing a black hole in the spacetime with dark vacuum energy represented by a cosmological constant. The detailed physical properties of such spacetimes have been treated in a series of works~\cite{Stuchlik83,Stuchlik99a,Stuchlik00a,Stuchlik02,Stuchlik04,Stuchlik05,Stuchlik08,Stuchlik2008arXiv,Stuchlik09a,Stuchlik16a,Charbulak17,Stuchlik18}. Here we are not interested in the explicit form of the solution for Eq. (\ref{rh}); however, we can see that the radius of horizon depends on two parameters, black hole charge $Q$ and $\lambda$ parameter and in the case when $\lambda=0$ horizon of the black hole will become $r_h=2M$, however for the nonzero value of $\lambda$ parameter, the situation will be quite complicated. Figure~\ref{fig:horizon} draws dependence of horizon radii from black hole charge for the different values of $\lambda$ parameter for GZ as well Reissner-Nordstr{\"o}m-de Sitter metrics. From Fig.~\ref{fig:horizon} one can see that there exists such a critical value of $\lambda$ parameter, i.e., $\lambda=\lambda_0$ that compensates the gravitating and antigravitating fields. On the other hand, the critical value $\lambda_0$ separates physical and nonphysical space. In the region, $\lambda<\lambda_0$ the gravitational field dominates, and in this region, we observe physical space, while in $\lambda>\lambda_0$ antigravitational field dominates which corresponds to nonphysical spacetime. In order to find the critical values of radius and $\lambda$ parameter, we use the following conditions $U=0$ and $U'=0$, and for GZ metric, we obtain
\begin{align}\label{r0}
&r_0=\frac{6M^2+Q^2\pm\sqrt{\left(Q^2-2M^2\right)\left(Q^2-18M^2\right)}}{4M}\ ,  \\\nonumber&\lambda_0=\frac{3\left(108M^4-36M^2Q^2-Q^4\right)}{16Q^6}\\&\pm\frac{3\left(Q^2-18 M^2\right) \sqrt{\left(Q^2-2M^2\right)\left(Q^2-18M^2\right)}}{16Q^6}\ ,
\end{align} 
while for Reissner-Nordstr{\"o}m-de Sitter metric one gets
\begin{align}\label{r00}
&r_0=\frac{1}{2}\left(3M\pm\sqrt{9M^2-8Q^2}\right)\ ,  \\\nonumber&\lambda_0=-\frac{3 \left(27 M^4-36 M^2Q^2+8Q^4\right)}{32Q^6}\\&\pm\frac{3M\sqrt{(9M^2-8Q^2)^3}}{32Q^6}\ .
\end{align} 
In the case, when $Q=0$, one gets $r_0=3M$ and $\lambda_0=1/9$. 
An interesting fact is the expressions for the critical value of radii in (\ref{r0}) and (\ref{r00}) provide the same expressions for the radii of photonsphere for both GZ and pure Reissner-Nordstr{\"o}m-de Sitter spacetime metrics. In next section, we will show detailed calculations of photonsphere radii.

\begin{figure*}[ht!]
\includegraphics[width=\hsize]{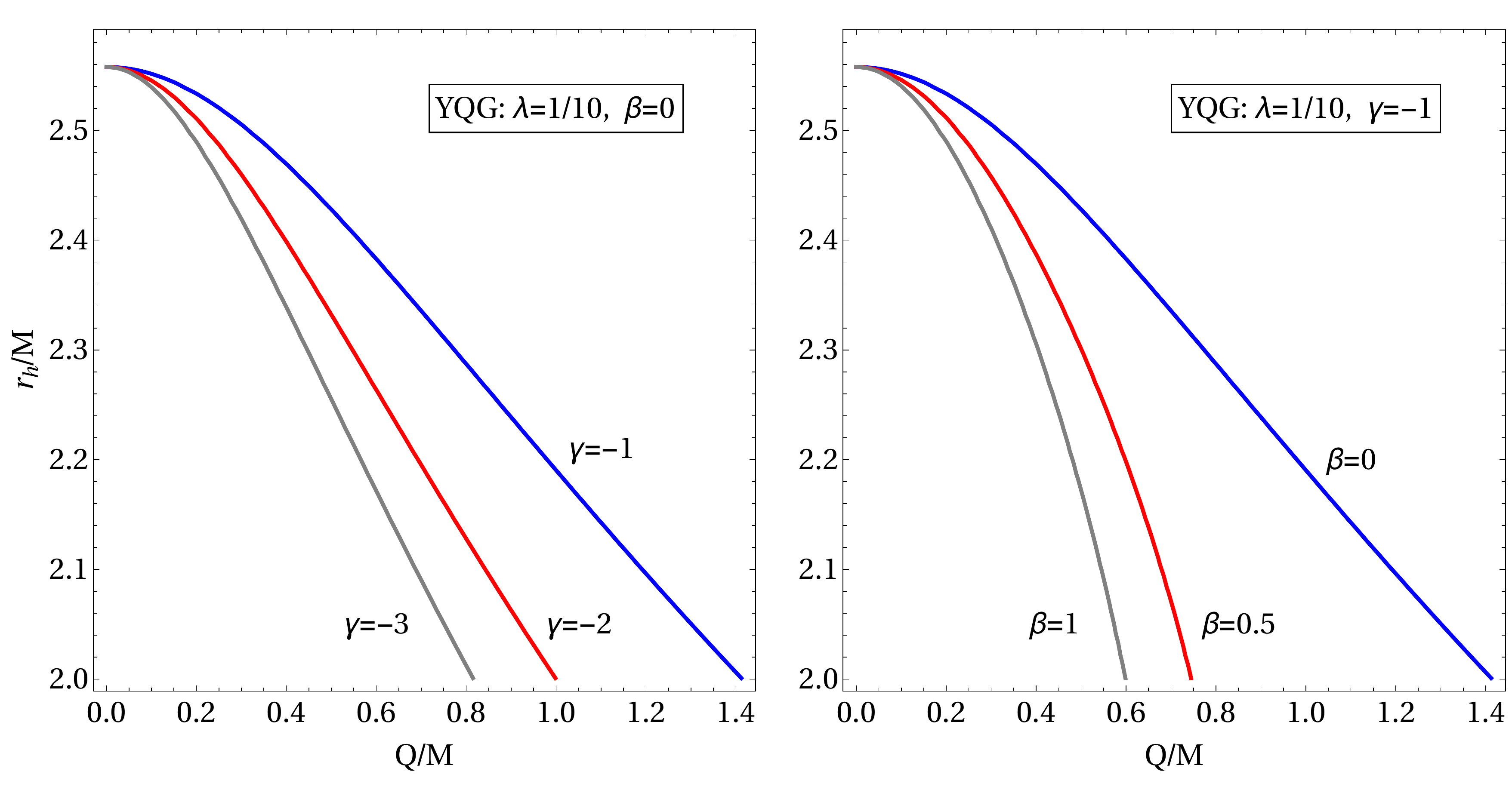}
\caption{(Color online) Left: dependence of the radius of the horizon from the black hole charge for different values of $\beta$ parameter. Right: dependence of the radius of the horizon from the black hole charge for different values of $\gamma$ parameter.\label{fig:Hor}}
\end{figure*}

From now we focus on the calculation of horizon radii in YQG solution and the exact form of the lapse function is given in Tab~\ref{Tab1}. In order to calculate the radii of the horizon, we again recall definition, $U=0$, which allows writing 
\begin{align}\label{rh1}
1-\frac{2M}{r}-\frac{\lambda}{3}r^2\left(1+\frac{\gamma Q^2}{Mr}\right)+\frac{\beta Q^2}{r^2\left(1+\frac{\gamma Q^2}{Mr}\right)}=0\ .    
\end{align}
From Eq.(\ref{rh1}), as we can see that the horizon's radius should depend on five different parameters, black hole mass $M$, charge $Q$, $\lambda$, $\gamma$, and $\beta$ parameters. Figure~\ref{fig:Hor} shows the dependence of the horizon radius from the black hole charge at $\lambda<\lambda_0$ region for the different values of $\gamma$ and $\beta$ parameters. Note that in $\lambda<\lambda_0$ there are two different radii of the horizon, the first one is the cosmological horizon which is not interesting and the second one is the physical horizon as shown in Fig.~\ref{fig:Hor}. One can also see that for various values of $\gamma$ and $\beta$ parameters the maximum value of the black hole charge decreases.
%%%%%%%%%%%%%%%%%%%%%%%%%%%%%%%%%%%%%%%%%%%%%%%%%%%%

\section{Geodesic motion\label{particlemotion}}

In this section, we study test particle motion around the black hole in the Einstein-Maxwell-scalar theory. Since the spacetime metric (\ref{metric}) is independent of coordinates $t$ and $\phi$, then corresponding momentum to these coordinates, $p_t$ and $p_\phi$ should be conserved, which are related to the energy and the angular momentum of a test particle at infinity. It is well known that the four-velocity of test particle, i.e., $u^\alpha=\dot x^\alpha=dx^\alpha/d\lambda$, where $\lambda$ is an affine parameter, is satisfied to the following normalization condition: $g_{\alpha\beta}u^\alpha u^\beta=-1$ which allows to write equation of motion in background geometry (\ref{metric}) in the form 
\begin{align}\label{PM}
\dot r^2+U(r) f(r) \dot\theta^2 = {\cal E}^2-V_{\rm eff}(r,\theta)\ ,     
\end{align}
where the effective potential $V_{\rm eff}(r,\theta)$ is defined as
\begin{align}\label{Veff}
V_{\rm eff}(r,\theta) = U(r)\left(1+\frac{{\cal L}^2}{f(r)\sin^2\theta}\right) \ ,
\end{align}
here ${\cal E}$ and ${\cal L}$ are, respectively, the specific energy and the specific angular momentum of a test particle at infinity. Notice that these quantities satisfy the following relations:
\begin{align}
{\cal E}=U(r)u^t\ , \quad {\cal L}=f(r)\sin^2\theta u^\varphi\ ,
\end{align}

Now we focus on the analysis of test particle motion by investigating the effective potential. The standard analysis shows that particle motion is limited by the energetic boundary conditions given by $\dot r=0$ and $\dot\theta=0$ which leads to
\begin{align}\label{V=0}
V_{\rm eff}(r,\theta)=0\ .    
\end{align}
On the other hand, the stationary points of the effective potential (\ref{Veff}) can be found from the following conditions
\begin{align}\label{stationary}
\partial_r V_{\rm eff}(r,\theta)=0\ ,\qquad \partial_\theta V_{\rm eff}(r,\theta)=0\ ,   
\end{align}
which allow to find extrema ($r_0, \theta_0$) of the function. From the second expression in (\ref{stationary}), one can easily find that one of the extremum points of the effective potential is located at $\theta_0=\pi/2$ which corresponds to an equatorial plane and another extremum point $r_0$ of the effective potential can be found from the first expression in (\ref{stationary}). Using the Eq.(\ref{V=0}) and the first expression in (\ref{stationary}), the radial dependence of the specific energy and the specific angular momentum at circular orbit can be found as
\begin{align}\label{energy}
{\cal E}^2=\frac{U^2 f'}{Uf'-fU'}\ ,
\\\label{momentum}
{\cal L}^2=\frac{f^2U'}{Uf'-fU'}\ .    
\end{align}

\begin{figure*}[ht!]
\includegraphics[width=\hsize]{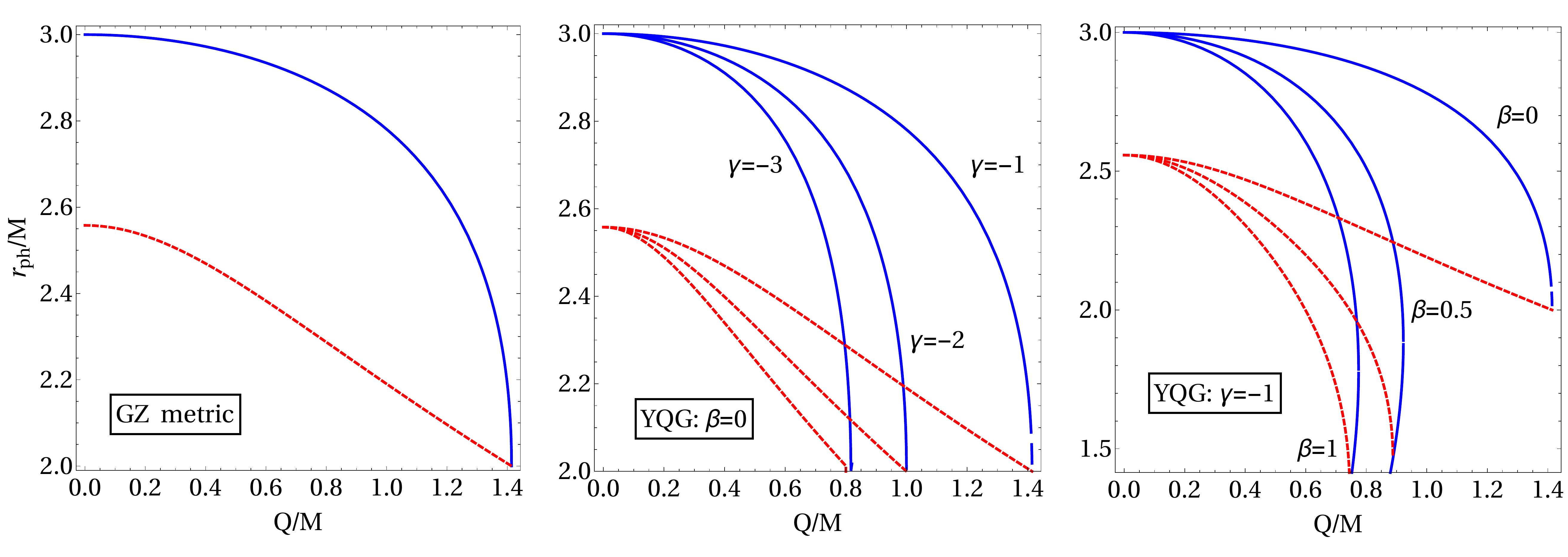}
\caption{(Color online) Left panel: dependence of the radius of photonsphere from black hole charge for both  GMGHS and GZ solutions. Central and Right panels: dependence of the radius of photonsphere from black hole charge for YQG metric for different values of $\beta$ parameter for the fixed value of $\gamma$ and the same plot for different values of $\gamma$ parameter at the fixed values of $\beta$ parameter. Solid (blue) line represents radius of photonsphere, while dashed line is responsible for radius of horizon at $\lambda=1/10$.\label{fig:photonsphere}}
\end{figure*}

One of the interesting features of black holes is related to characteristic radii around them, from these points of view we are now interested in the determination of the characteristic radii around rotating black holes, namely, photonsphere, marginally bound, and ISCO radii. We first focus on the determination of the radius of the photonsphere. In fact that photon is massless, i.e., $m=0$, which means its specific energy tends to infinity (i. e.  ${\cal E}\to \infty$). In order to find the radii of photonphere, we use the fact that denominator of the expression (\ref{energy}) be zero, i.e., $Uf'-fU'=0$. Our calculations show that the expression for the radius of photonsphere in GMGHS and GZ solutions is exactly the same and has a form
\begin{align}\label{PH}
r_{\rm ph}^\pm=\frac{6M^2+Q^2\pm\sqrt{\left(Q^2-2M^2\right)\left(Q^2-18M^2\right)}}{4M}\ ,    
\end{align} 
which is independent of $\lambda$ parameter and $r_+$ is responsible for the radius of the photonsphere in GMGHS and GZ solutions. In order to calculate the radius of photonsphere for YQG metric, one has to solve the following cubic equation: 
\begin{align}\label{cubic}
x^3+px+q=0\ ,    
\end{align}
where 
\begin{align}
&x=\frac{r}{M}-1+\frac{\gamma Q^2}{2M^2}\ ,
\\&p=-3-\frac{2(\gamma-\beta)Q^2}{M^2}-\frac{\gamma ^2 Q^4}{4 M^4}\ ,
\\&q=-2-\frac{2(\gamma-\beta)Q^2}{M^2}-\frac{\gamma^2 Q^4}{2 M^4}\ .
\end{align}
The cubic equation has three different solutions as
\begin{align}
x_k=2\sqrt{-\frac{p}{3}}\cos\left[\frac{1}{3}\arccos\left(\frac{3q}{2p}\sqrt{\frac{-3}{p}}\right)-\frac{2\pi k}{3}\right] \ ,    
\end{align}
where $ k=0,1,2$. Then the radius of the photonsphere can be expressed as $r_k=M(x_k+1+\gamma Q^2/2M^2)$. In Fig.~\ref{fig:photonsphere}, the various dependence of the radius of the photon sphere is illustrated. From Fig.~\ref{fig:photonsphere}, one can see that with increasing $Q$ parameter the radius of the photonsphere decreases. Other interesting fact is that the critical value of the black hole charge decreases with various values of $\gamma$ and $\beta$ parameters of the black hole and they also independent of $\lambda$ parameter. 

\begin{figure*}[ht!]
\includegraphics[width=\hsize]{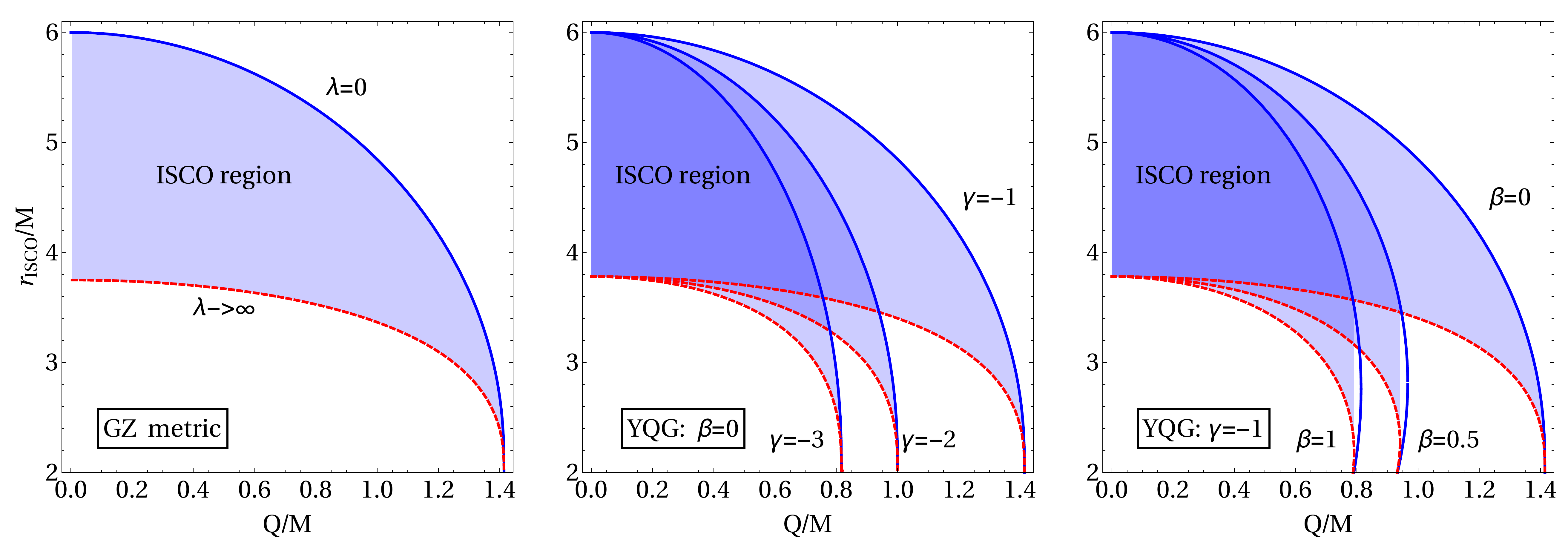}
\caption{(Color online) Left panel: dependence of ISCO radius from black hole charge for GZ solution. Shaded region represents ISCO radii for any value of $\lambda$ parameter. Upper solid (blue) line represents ISCO radius in GMGHS spacetime. Central and right panels: dependence of ISCO radii from black hole charge in YQG metric for the different values of $\beta$ parameter for fixed value $\gamma$ and the same plot is for different values of $\gamma$ parameters at the fixed values of $\beta$ parameter. The lower limit of ISCO radii equal to $r/M=15/4$ at $Q=0$.\label{fig:ISCO}}
\end{figure*}

There exists other radius of massive particle, the so-called innermost stable circular orbit (ISCO), which can be found using the following conditions $\partial_r^2 V_{\rm eff}(r,\pi/2)=0$ along with Eqs.(\ref{V=0}) and (\ref{stationary}). Hereafter simple algebraic calculation, one can get 
\begin{align}\label{ISCO}
\frac{2Uf'U'}{f}-\frac{Uf''U'}{f'}+UU''-2 U'^2=0\ . 
\end{align}
The ISCO radius for a test particle can be found as the solution of the equation (\ref{ISCO}) being all three black hole solutions. One has to emphasize that hereafter making normalizing the radial coordinate with the black hole mass, i.e., $r/M$, then ISCO radius will be dependent on the black hole charge $Q$, the cosmological parameter $\lambda$, and also $\beta$, $\gamma$ parameters. Figure~\ref{fig:ISCO} draws dependence of the ISCO radius of the test particle from a black hole charge for different sets of $\beta$ and $\gamma$ parameters. The left panel of Fig.~\ref{fig:ISCO} shows the dependence of ISCO radius from the black hole charge in GZ metric for arbitrary values of the cosmological parameter $\lambda$. It is interesting to mention that for zero value of the cosmological parameter ISCO radius represents the upper limit given with blue solid line in Fig.~\ref{fig:ISCO}. The lower limit of the ISCO curve (dashed red line) is not changed after some value of the cosmological parameter. We have found that for any value of the cosmological parameter the ISCO radius will be $r/M=15/4$ at $Q=0$. One can see that both upper and lower ISCO radii decrease up to the horizon radius. In the central and right panels of Fig.~\ref{fig:ISCO} same plots together with left one are presented for different sets of the other $\beta$ and $\gamma$ parameters in YQG metric.

\subsection{Black holes in EMS theory versus rotating Kerr black holes }

It is well-known that astrophysical black holes can be described by a few parameters like mass, spin, and/or (electric and magnetic) charge. However, it is difficult to measure these parameters directly in the astrophysical observations except the total mass which can be used by the Newtonian dynamics. One way is to estimate the values of the black hole parameters through measurements of indirect observations (for example dynamics of photons and massive particles). In other words, one can measure and/or compare their effects on the particle dynamics by considering two-parameters model for the central black hole: mass-spin and mass-charge. On the other hand, in fact, that the effect of spin of the rotating Kerr black hole and electric charge of a static black hole is similar to the ISCO radius. The question arises from this degeneracy, how one can distinguish the effects of spin and charge parameters based on the observational data from dynamics of the particle around a black hole? Moreover, even in the charged static black hole model, it is important to know the type of charge. In this subsection, we will try to show the cases in which values of spin and charge parameters ISCO radius of test particles around rotating Kerr and charged GZ/YQG static black holes will be exactly the same. 

The expression for ISCO radius of test particles corresponding to retrograde and prograde orbits around rotating Kerr black hole can be found in~\cite{Bardeen72}
\begin{align}\label{iscoeqKerr}
r_{\rm isco}= 3 + Z_2 \pm \sqrt{(3- Z_1)(3+ Z_1 +2 Z_2 )} \ ,
\end{align}
where denotes,
\begin{align} \nonumber
Z_1 & =  
1+\left( \sqrt[3]{1+a}+ \sqrt[3]{1-a} \right) 
\sqrt[3]{1-a^2} \ ,
\\ \nonumber
Z_2 & =  \sqrt{3 a^2 + Z_1^2} \ .
\end{align}
Here we will provide analysis of degeneracy values of spin parameter and charge of black holes in EMS theories using Eqs. (\ref{ISCO}) and (\ref{iscoeqKerr}). 

\begin{figure*}[ht!]
\includegraphics[width=0.4\linewidth]{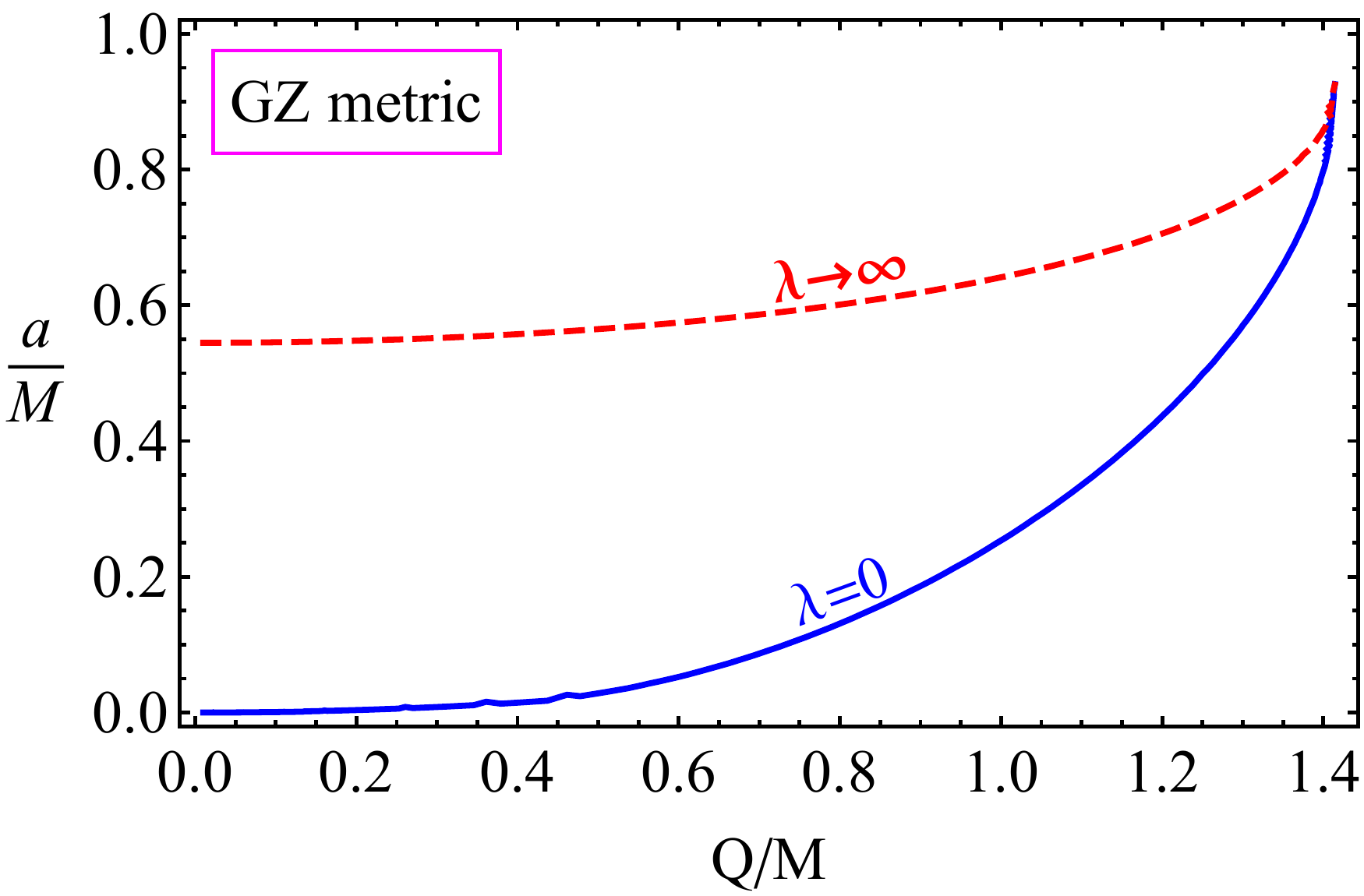}
\includegraphics[width=0.4\linewidth]{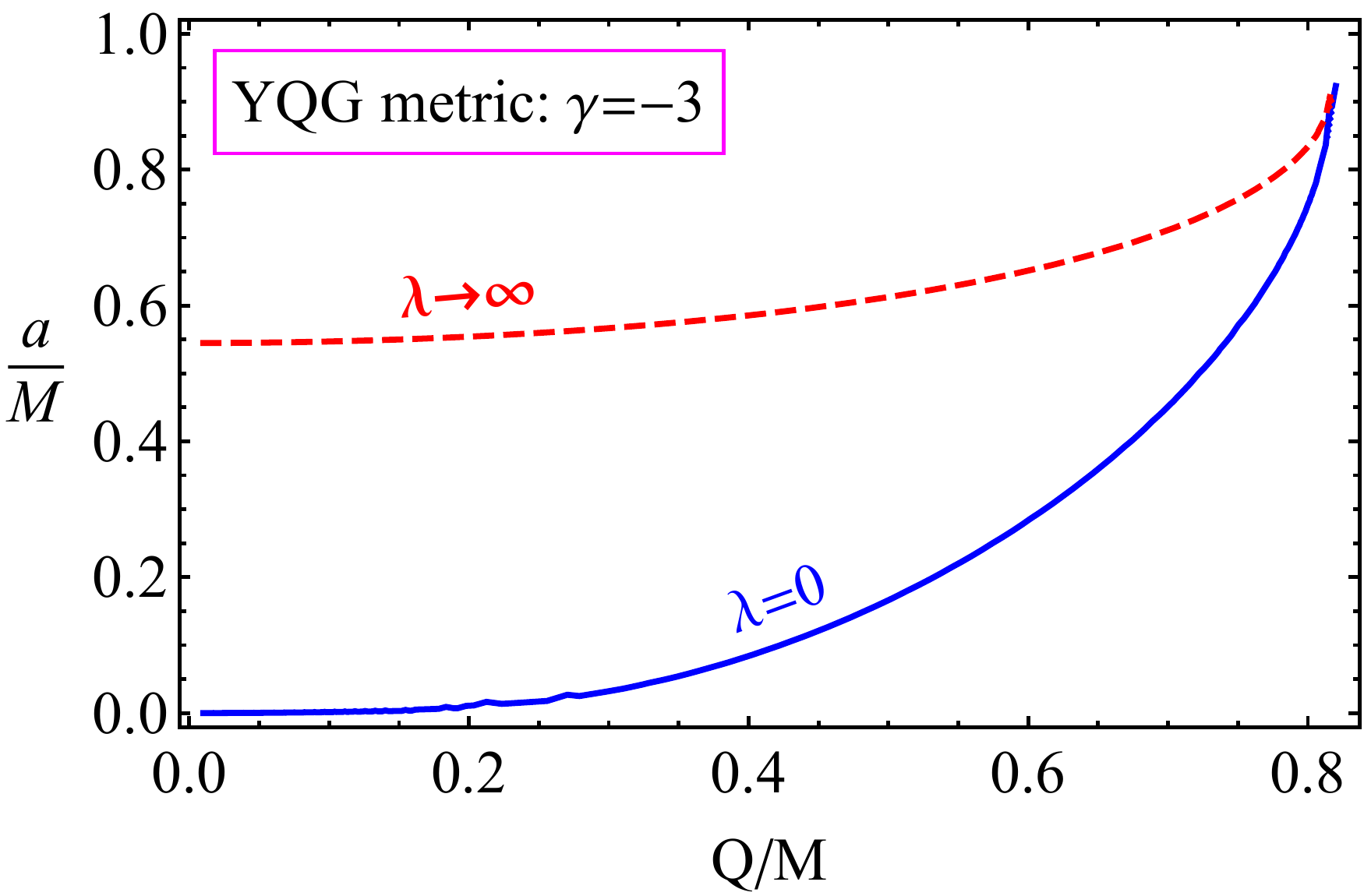}
\caption{Relations of degeneracy values of charge of GZ (left panel) and YQG (right panel) black holes and spin of Kerr black hole for the same ISCO radius. 
\label{fig:avsQYQG}}
\end{figure*}

One can see from Fig.~\ref{fig:avsQYQG}  that GZ black hole charge can mimic up to $a/M=0.925877$ providing the same value for ISCO radius of test particles for the case when $\lambda=0$, while at $\lambda \to \infty$ the charge can mimic the spin parameter in the range of $a/M \in (0.544736, \ 0.925877)$ (see the red dashed line). Moreover, the same mimic values for YQG black hole charge, but in this case, the maximum value for the charge is $Q/M \simeq 0.81$. In realistic astrophysical applications one may estimate the charge of the supermassive GZ black hole Sgr A* (M87) as $Q/M=1.203$ ($Q/M=1.41392$) when the parameter $\lambda=0$, while charge of the SMBH YQG Sgr A* is $Q/M=0.694553$ ($Q/M=0.848621$) when parameters $\lambda=0$ and $\beta=-3$.

%%%%%%%%%%%%%%%%%%%%%%%%%%%%%%%%%%%%%%%%%%%%

\section{Fundamental frequencies}

\begin{figure*}
\includegraphics[width=0.8\textwidth]{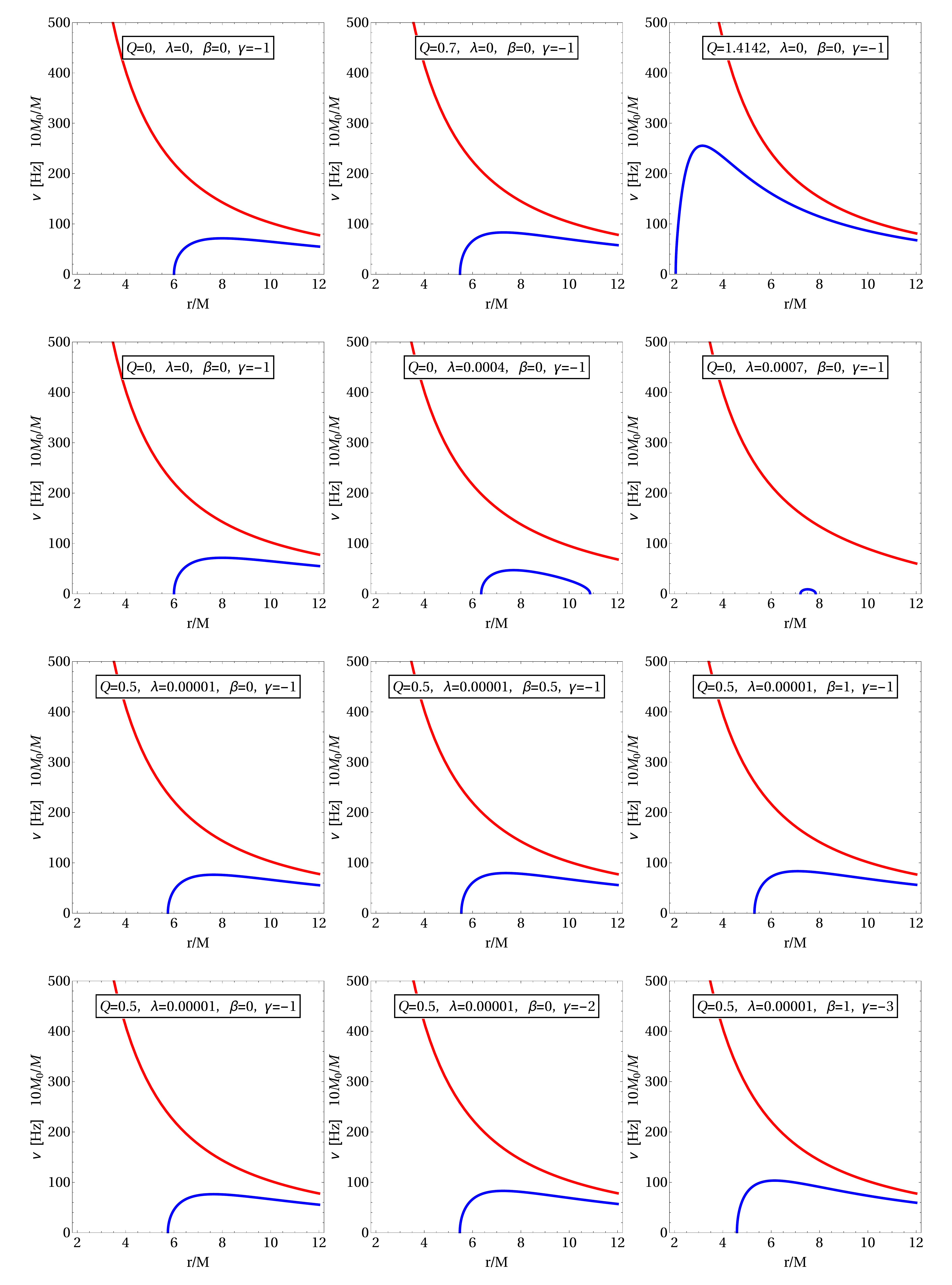}
\caption{(Color online) The radial dependence of the fundamental frequencies governed by test particle motion around the black hole.\label{fig:QPO}}
\end{figure*}

In this section, we will show in detail the derivation of expression for the fundamental frequencies governed by test particle orbiting around the black hole in the Einstein-Maxwell-scalar theory which is one of the simple models to explain Quasi-Periodic oscillations (QPO) observable around compact astrophysical objects. The angular velocity measured by a distant observer can be found as
\begin{align}
\Omega=\sqrt{-\frac{\partial_r g_{tt}}{\partial_rg_{\phi\phi}}}=\sqrt{\frac{U'(r)}{f'(r)}}\ .
\end{align}
However, there might exist other types of oscillations characterized by the radial and vertical frequencies. These radial and vertical fundamental frequencies can be calculated by considering small perturbation around the circular orbit, respectively, along with the radial $r\to r_0+\delta r$ and azimuthal $\theta\to \theta_0+\delta\theta$ directions. Then the effective potential can be expanded in terms of $r$ and $\theta$ coordinates in the form
\begin{align}\label{Vexpand}
\nonumber
&V_{\rm eff}(r,\theta)=V_{\rm eff}(r_0,\theta_0)
\\\nonumber
&+\delta r\,\partial_r V_{\rm eff}(r,\theta)\Big|_{r_0,\theta_0} +\delta\theta\, \partial_\theta V_{\rm eff}(r,\theta)\Big|_{r_0,\theta_0}
\\\nonumber
&+\frac{1}{2}\delta r^2\,\partial_r^2 V_{\rm eff}(r,\theta)\Big|_{r_0,\theta_0}+\frac{1}{2}\delta\theta^2\,\partial_\theta^2 V_{\rm eff}(r,\theta)\Big|_{r_0,\theta_0}
\\
&+\delta r\,\delta\theta\,\partial_r\partial_\theta V_{\rm eff}(r,\theta)\Big|_{r_0,\theta_0}+{\cal O}\left(\delta r^3,\delta\theta^3\right)\ .
\end{align}
The careful analysis of this expansion shows that the first term of (\ref{Vexpand}) vanishes due to the condition (\ref{V=0}), on the other hand, using the stability conditions of the effective potential in Eq.(\ref{stationary}) one can remove the second, third and last terms of the expression (\ref{Vexpand}), which means the only two terms remain which are proportional to the second-order derivatives from the effective potential with respect to $r$ and $\theta$. This is the standard way of the derivation of the equation of motion for the harmonic oscillator. Now we focus on equation of motion once again and before continuing our calculations we replace a derivation with respect to an affine parameter in Eq.(\ref{PM}) into the time derivation (i.e., $dt/d\lambda=u^t$); the main idea of this replacement is to express all equations in terms of physical quantities measured by a distant observer. Now substituting the expression (\ref{Vexpand}) into (\ref{PM}) and following the above statements, we obtain harmonic oscillator equations (Euler-Lagrange equations) for displacements $\delta r$ and $\delta\theta$ in the form
\begin{align}
\frac{d^2\delta r}{dt^2}+\Omega_r^2 \delta r=0\ , \qquad \frac{d^2\delta\theta}{dt^2}+\Omega_\theta^2 \delta\theta=0\ ,   
\end{align}
where $\Omega_r$ and $\Omega_\theta$ are, respectively, the radial and azimuthal angular frequencies measured by a distant observer, defined as
\begin{align}
&\Omega_r^2=\frac{1}{2(u^t)^2}\partial_r^2V_{\rm eff}(r,\theta)\Big |_{\theta=\pi/2}\ ,
\\
&\Omega_\theta^2=\frac{1}{2fU(u^t)^2}\partial_\theta^2V_{\rm eff}(r,\theta)\Big |_{\theta=\pi/2}\ .
\end{align}
Finally, expressions for the fundamental frequencies in the spacetime of the black hole in EMS theory governed by test particle take the form
\begin{align}\label{Omegar}
&\Omega_r=\Omega\sqrt{U\left(\frac{f'U''}{2U'}+\frac{f'^2}{f}-1\right)-f'U'}\ ,
\\\label{Omegat}
&\Omega_\theta=\Omega=\sqrt{\frac{U'}{f'}}\ ,
\\\label{Omegaf}
&\Omega_\phi = \Omega =\sqrt{\frac{U'}{f'}}\ .
\end{align}
Here we can see that the expressions for both vertical and
azimuthal angular velocities are same which means we
cannot distinguish them when we observe them. Notice that the expressions (\ref{Omegar})-(\ref{Omegaf}) represent the angular frequencies; however, the fundamental frequencies can be expressed as $\nu_i=\Omega/(2\pi)$, then the Keplerian frequency takes a form:  
\begin{equation}
\nu = \frac{1}{2\pi}\frac{c^3}{GM}\left(\frac{GM}{c^2r}\right)^{3/2}{\rm Hz}\ .
\end{equation}
In order to have an idea to understand the theoretical expressions for fundamental frequencies one can write the expression for the Keplerian frequency for the stellar black hole in the Schwarzschild spacetime in the form:
\begin{equation}
\nu \simeq 220\left(\frac{M}{10M_\odot}\right)^{-1}\left(\frac{6GM}{c^2r}\right)^{3/2}{\rm Hz}\ .
\end{equation}
Notice that the fundamental frequencies are observed only outside the region of the ISCO radii of the test particle. Similarly, the expressions for the fundamental frequencies can be obtained around black holes in EMS theory. The radial dependence of the fundamental frequencies governed by test particle motion around the black hole for the different sets of the black hole parameters is illustrated in Fig.~\ref{fig:QPO}. Most of the observations show that fractional frequencies to be order of $\nu_\phi:\nu_r=3:2$, $\nu_\theta:\nu_r=3:2$ or these ratio might be combinations of fundamental frequencies. Of course, here in Fig.~\ref{fig:QPO}, it is difficult to analyse these results for the full set of the black hole parameters and compare them with observational data. However, we can see (the second row of Fig.~\ref{fig:QPO}) that due to the cosmological parameter ratio of the frequencies gets larger which disagrees with observations. The $\nu_{\theta,\phi}:\nu_r=3:2$ ratio can obtained with playing of all other black hole parameters.

%%%%%%%%%%%%%%%%%%%%%%%%%%%%%%%%%%%%%%%%%%%%%%%%%%

\section{Charged particle motion}

In this section, we study charged particle motion around a charged black hole in EMS theory. The equation for the radial motion for the charged particle is
\begin{align}
\dot r^2 = \left[{\cal E}+eA_t(r)\right]^2+U(r)\left[1+\frac{{\cal L}^2}{f(r)}\right]\ ,
\end{align}
where $e$ is charge of particle. The radial functions $f(r)$, $U(r)$, and $A_t(r)$ are shown in Table~\ref{Tab1} for given parameters. The effective potential $V_{\rm eff}(r)$ for charged particle can be found from the following conditions $\dot r=0$ and ${\cal E}=V_{\rm eff}(r)$, in the form
\begin{align}\label{Effpot}
V_{\rm eff}(r)=\sqrt{U(r)\left[1+\frac{{\cal L}^2}{f(r)}\right]}-eA_t(r)\ ,
\end{align}
where the first term of the effective potential is responsible for pure neutral particle case, while the second one represents the Coulomb interaction of the charged particle with a charged black hole. In the absence of particle charge, i.e., $e=0$, the effective potential reduces to the one for neutral particle.

\begin{figure}
\centering
\includegraphics[width=0.48\textwidth]{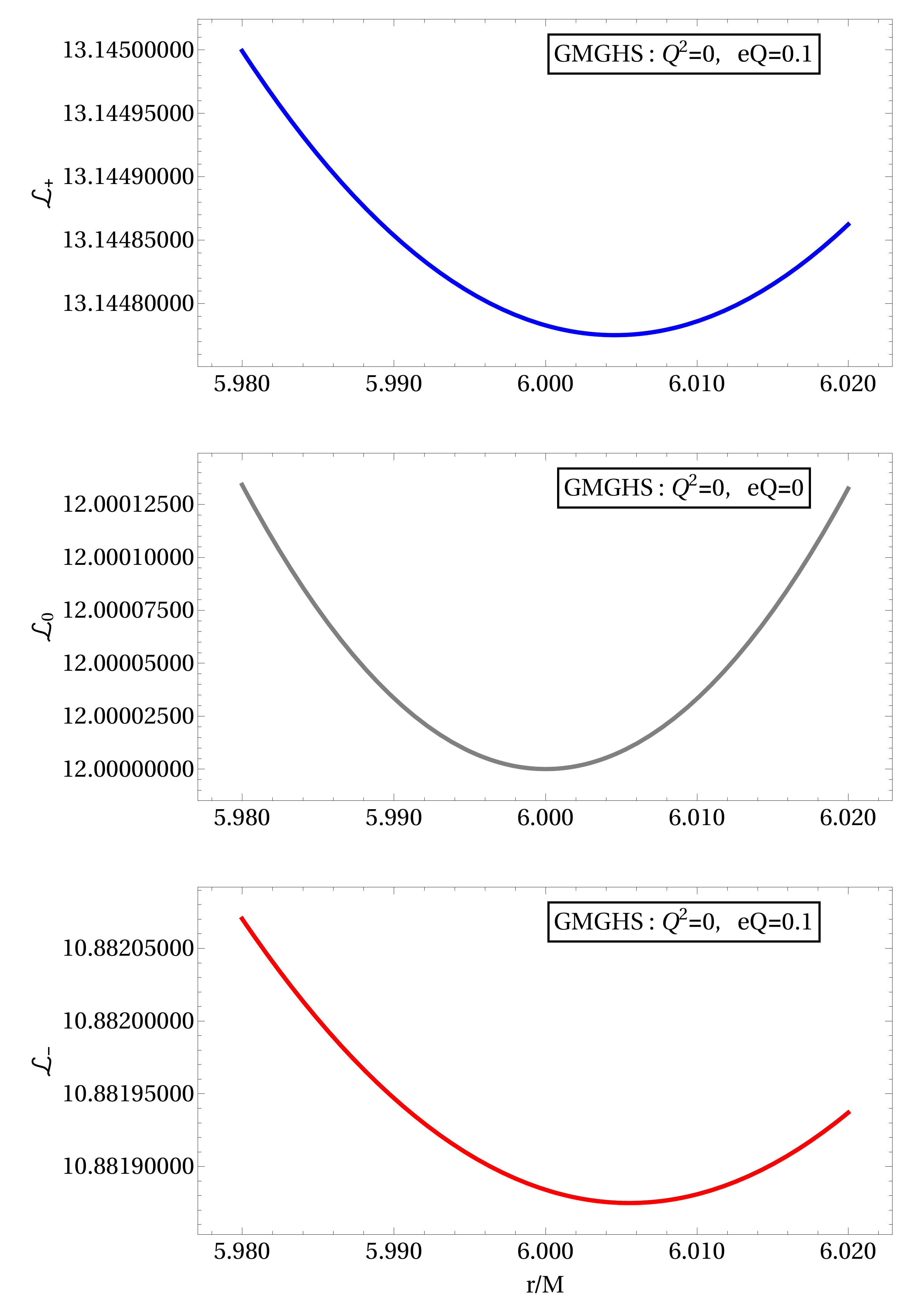}
\caption{(Color online) Radial dependence of the specific angular momentum ${\cal L}_\pm$ (Top and Bottom panels) of test particle for the fixed values of the coupling parameter $eQ$ and (Central panel) for neutral particle ${\cal L}_0$.}
\label{fig:ChargedMoment}
\end{figure}

Now we concentrate on finding ISCO radius for the charged particles around the charged black hole in EMS theory. We have already analysed behavior of ISCO radii for the neutral particle in EMS theory. A more precise way of showing this argument is to follow again to the standard procedure, in which $V_{\rm eff}={\cal E}^2$, $V_{\rm eff}'(r)=0$ which allows finding expressions for the extrema of the specific angular momentum and the specific energy. For the moment, we focus on the specific angular momentum of charged particle,
\begin{align}\nonumber
{\cal L}_{\pm}^2&=\frac{f^2U'}{Uf'-fU'}+\frac{2f^3U(eA_t') 2}{\left(Uf'-fU'\right)^2}\\&\pm\frac{2f^2UeA_t'\sqrt{f 2(eA_t')^2+f'(Uf'-fU')}}{\left(Uf'-fU'\right)^2}\ .
\label{Lpm}
\end{align}
which reduces ${\cal L}=f^2U'/(Uf'-fU')$ for neutral particle as shown in Eq.(\ref{momentum}). From Eq.(\ref{Lpm}) one can see that the specific angular momentum of the charged particle ${\cal L}_\pm$ satisfies the following symmetry:
\begin{align}
{\cal L}_+(eQ)={\cal L}_-(-eQ)\,\quad \text{\rm or}\quad {\cal L}_+(-eQ)={\cal L}_-(eQ)\ .
\end{align}
Now one can find the minimum of ${\cal L}_\pm$ in Eq.(\ref{Lpm}) which corresponds to the ISCO radius of charged particle and see how it changes for different values of the coupling parameter $eQ$. In order to make a qualitative analysis of the dependence of ISCO radius $r_{\rm ISCO}$ for charged particle from the coupling parameter $eQ$, we assume that square of the black hole charge is negligibly small (i.e., $Q^2\to 0$) to change spacetime geometry and keeping the interaction term in Eq.(\ref{Lpm}) one can check the radial dependence of the specific angular momentum ${\cal L}_\pm$ of the charged particle. Figure~\ref{fig:ChargedMoment} shows the radial dependence of specific angular momentum of a charged particles in GMGHS for the fixed values of the coupling parameter. Here one has to emphasize that extremum of the specific angular momentum is responsible for to ISCO radius. As one can see from Fig.~\ref{fig:ChargedMoment} the minimum of the specific angular momentum for charged particle cases shifts into large values of the radial coordinate, which concludes that ISCO radius increases for both positively and negatively charged particles. The ISCO radius increases for charged particles even when $Q = 0$.
%Note that the result in Fig.~\ref{fig:ChargedMoment} will be the same in RN spacetime.  Keep in mind that here we took $Q 2=0$, however restoring black hole charge in the [removed]\ref{Lpm}) one can show that ISCO radii for charged particle increase any cases.
%

\begin{figure}
\centering
\includegraphics[width=0.48\textwidth]{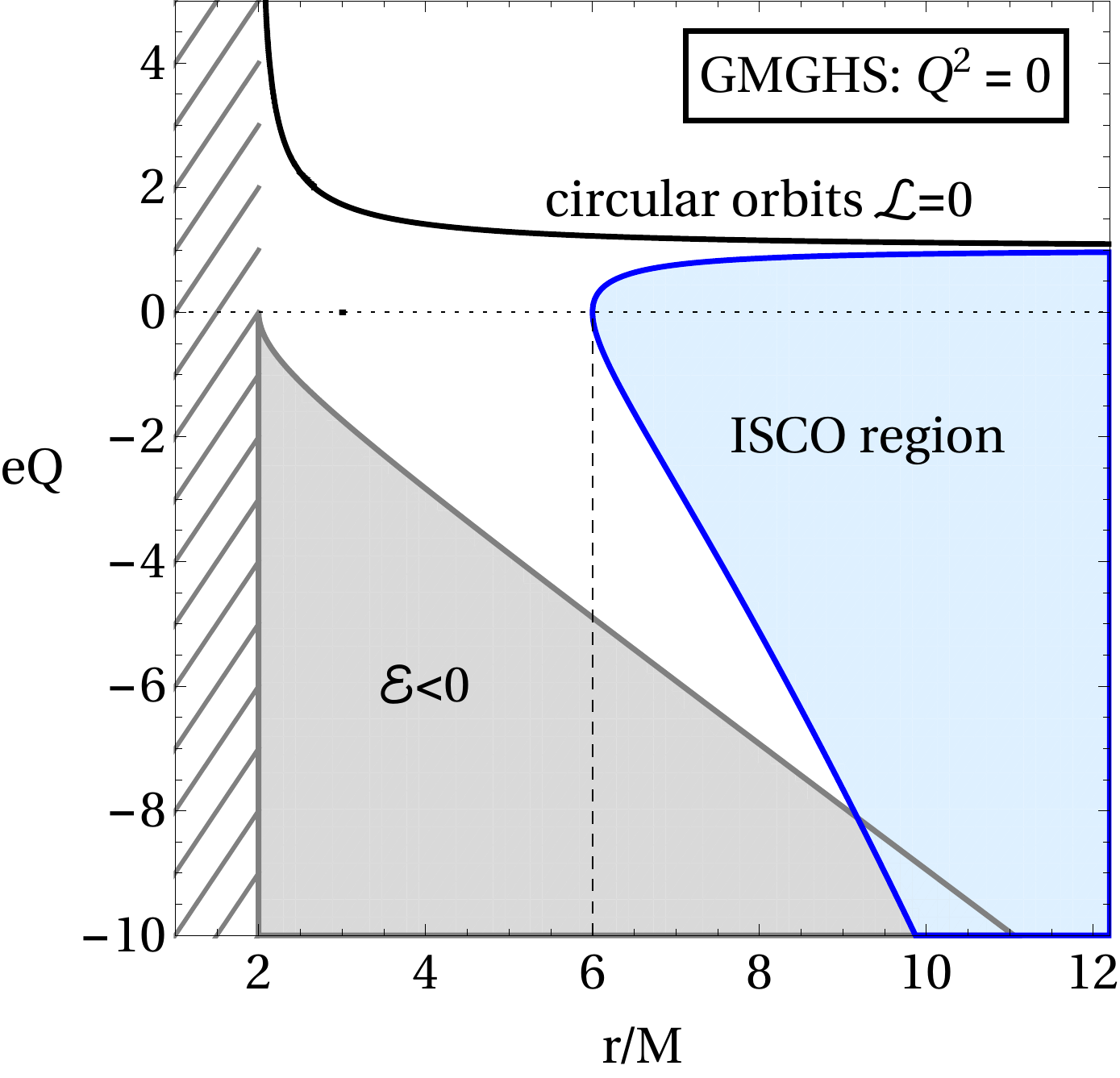}
\caption{(Color online) ISCO radius for charged particle is a function of the coupling parameter $eQ$.\label{fig:ChargedISCO}}
\end{figure}

Now we focus on the definition of the ISCO radius for charged particles and see how it depends on the coupling parameter. In order to obtain the ISCO radius for the charged particle, we use $V_{\rm eff}''(r)=0$ and eliminating ${\cal L}_{\pm}$ by using (\ref{Lpm}) we obtain a long equation for radial coordinate. Hereafter making careful numerical analysis on obtained equation we present dependence of ISCO radius for a charged particle for different values of the black hole charge $Q$ and coupling parameter $eQ$. Figure~\ref{fig:ChargedISCO} illustrates the dependence of ISCO radius from the coupling parameter $eQ$. The light blue region corresponds to the stable circular orbits, one can see that the lower limit of the blue line is $6M$ for zero value of the coupling parameter and with increasing the coupling parameter $qQ$. ISCO radius increases in both positively and negatively charged particles. The black line in Fig.~\ref{fig:ChargedISCO} is responsible for the circular orbit of a charged particle when angular momentum to be taken a zero, while the gray region represents the negative energy of the charged particle.

As we have already mentioned above, our aim is to investigate the charged particle motion around the charged black hole to see the significance of the coupling parameter in the ISCO radius for a charged particle in the following approximation, $Q^2\to 0$. So far, we have studied this issue in GMGHS metric only, and our calculations show that ISCO radii increase due to only the coupling parameter in all three different black hole solutions. However, it decreases due to the black hole's own parameters such as black hole charge $Q$, $\lambda$, $\beta$, and $\gamma$ parameters as we have shown in the case of neutral particle motion in the previous section. For nonzero values of black hole parameters, we will have very similar results as shown in Fig.~\ref{fig:ChargedISCO}, and the only difference is that the (blue) ISCO region will shift into the left side depending on the type of the black hole parameters variation.

%%%%%%%%%%%%%%%%%%%%%%%%%%%%%%%%%%%%%%%%%%%%%%%%%%%%

\section{Conclusion\label{Summary}}

In the present paper, we have tested three different GMGHS, GZ, and YQG black hole solutions that belong to EMS theory by considering test neutral and charged particle motion around the black hole. Obtained results can be summarized as follows:

\begin{itemize}
\item 
In order to investigate the spacetime properties of three different black hole solutions, we first determined the curvature invariants such as Ricci scalar, Ricci square, and the Kretschmann scalar. Our calculations have shown that all three metrics are characterized by two singularities, the first one is at $r=0$ being similar to other spherical-symmetric solutions such as in Schwarzschild, Reissner-Nordstr{\"o}m, etc. and other one arises at $r=Q^2/M$ ($r=-\gamma Q^2/M$ in YQG spacetime metric) which corresponds to a naked singularity.

\item
We have investigated the structure of the horizon of the black hole in three different solutions. It has been shown that there are two different regions that can be divided by the critical value of the cosmological parameter $\lambda_0$. In the region, $\lambda<\lambda_0$ horizon exists, while when $\lambda>\lambda_0$ horizon can not be observed. It has been found that the maximal value of the black hole charge is $Q_{\rm max}\leq \sqrt{2}M$.      

\item
The photonsphere around the black hole in EMS theory has also been investigated. It is found that in all three solutions the radius of the photonsphere does not depend on the cosmological parameter $\lambda$ and equal to $3M$ when $Q=0$ like in Schwarzschild spacetime. It decreases when the black hole charge gets larger and reaches up to $2M$ at the maximum value of the black hole where the horizon's radius takes the same value.

\item
We have studied the ISCO radius for a neutral particle orbiting around the black hole in EMS theory. It is shown that in all solutions ISCO radius for neutral particle decreases when the black hole charge gets large value. In the absence of the black hole charge and cosmological constant (i.e., $Q=0$, $\lambda=0$) in GZ solution, ISCO radius will become $r_{\rm ISCO}=6M$, while in the case when $Q=0$, $\lambda\neq 0$, ISCO radius takes the form $r_{\rm ISCO}/M=15/4$. Minimal value of the stable orbit takes $2M$ for the maximum value of the black hole charge.   

\item
We have also studied charged particle motion around the black hole in EMS theory. In this case, the charged particle is considered in the presence of the gravitational field and the Coulomb potential, which can be either attractive or repulsive depending on the interaction parameter $eQ>0$ or $eQ<0$. It is shown that in both cases, the ISCO radius for charged particles increases depending on the selected values of the coupling parameter.

\item
We have determined the fundamental frequencies governed by test particle orbiting around the black hole in EMS theory. We have presented the general form of the exact expressions for the fundamental frequencies in all three solutions. We have shown that the expressions for the angular part of frequencies to be equal to each other $\nu_\theta=\nu_\phi$. Finally, we have produced radial dependence of the fundamental frequencies that can be useful to fit with observational data. We have shown that it is impossible to get $\nu_\theta:\nu_r=3:2$ fractional frequencies in the presence of the cosmological parameter.  

\item

Finally, we have investigated the finding of the mimic values of charge parameter in GZ/YQG black hole solution which provides the same value for ISCO radius of test particles in the Kerr black hole. Detailed analysis showed that the charge of GZ (YQG) black hole can mimic the spin of Kerr black hole up to value of the spin parameter $a/M=0.925877$ when $\lambda=0$, while at $\lambda \to \infty$ the charge can mimic the spin parameter in the range of $a/M \in (0.544736, \ 0.925877)$. Moreover, the same mimic values for YQG black hole charge, but in this case of the maximum value for the charge is $Q/M \simeq 0.81$. In order to make astrophysical applications we have rough estimation on the charge of the GZ black hole Sgr A* (M87) as $Q/M=1.203$ ($Q/M=1.41392$) when the parameter $\lambda=0$, while charge of the SMBH YQG Sgr A* is $Q/M=0.694553$ ($Q/M=0.848621$) when parameters $\lambda=0$ and $\beta=-3$.

\end{itemize}

\section*{Acknowledgments}

This research is supported by Grants No. VA-FA-F-2-008, 
No.MRB-AN-2019-29 of the Uzbekistan Ministry for Innovative Development. 
J.R., A.A., and B.A. thank Silesian University in Opava for the 
hospitality during their visit. A.A. is supported by PIFI fund 
of Chinese Academy of Sciences. B.T. acknowledges the support %
from the internal student Grant No. SGS/12/2019  of SU.

\bibliographystyle{apsrev4-1}  
\bibliography{gravreferences}

\begin{thebibliography}{112}%
\makeatletter
\providecommand \@ifxundefined [1]{%
 \@ifx{#1\undefined}
}%
\providecommand \@ifnum [1]{%
 \ifnum #1\expandafter \@firstoftwo
 \else \expandafter \@secondoftwo
 \fi
}%
\providecommand \@ifx [1]{%
 \ifx #1\expandafter \@firstoftwo
 \else \expandafter \@secondoftwo
 \fi
}%
\providecommand \natexlab [1]{#1}%
\providecommand \enquote  [1]{``#1''}%
\providecommand \bibnamefont  [1]{#1}%
\providecommand \bibfnamefont [1]{#1}%
\providecommand \citenamefont [1]{#1}%
\providecommand \href@noop [0]{\@secondoftwo}%
\providecommand \href [0]{\begingroup \@sanitize@url \@href}%
\providecommand \@href[1]{\@@startlink{#1}\@@href}%
\providecommand \@@href[1]{\endgroup#1\@@endlink}%
\providecommand \@sanitize@url [0]{\catcode `\\12\catcode `\$12\catcode
  `\&12\catcode `\#12\catcode `\^12\catcode `\_12\catcode `\%12\relax}%
\providecommand \@@startlink[1]{}%
\providecommand \@@endlink[0]{}%
\providecommand \url  [0]{\begingroup\@sanitize@url \@url }%
\providecommand \@url [1]{\endgroup\@href {#1}{\urlprefix }}%
\providecommand \urlprefix  [0]{URL }%
\providecommand \Eprint [0]{\href }%
\providecommand \doibase [0]{http://dx.doi.org/}%
\providecommand \selectlanguage [0]{\@gobble}%
\providecommand \bibinfo  [0]{\@secondoftwo}%
\providecommand \bibfield  [0]{\@secondoftwo}%
\providecommand \translation [1]{[#1]}%
\providecommand \BibitemOpen [0]{}%
\providecommand \bibitemStop [0]{}%
\providecommand \bibitemNoStop [0]{.\EOS\space}%
\providecommand \EOS [0]{\spacefactor3000\relax}%
\providecommand \BibitemShut  [1]{\csname bibitem#1\endcsname}%
\let\auto@bib@innerbib\@empty
%</preamble>
\bibitem [{\citenamefont {{Gibbons}}\ and\ \citenamefont
  {{Maeda}}(1988)}]{Gibbons88}%
  \BibitemOpen
  \bibfield  {author} {\bibinfo {author} {\bibfnamefont {G.~W.}\ \bibnamefont
  {{Gibbons}}}\ and\ \bibinfo {author} {\bibfnamefont {K.-I.}\ \bibnamefont
  {{Maeda}}},\ }\href {\doibase 10.1016/0550-3213(88)90006-5} {\bibfield
  {journal} {\bibinfo  {journal} {Nuclear Physics B}\ }\textbf {\bibinfo
  {volume} {298}},\ \bibinfo {pages} {741} (\bibinfo {year}
  {1988})}\BibitemShut {NoStop}%
\bibitem [{\citenamefont {{Garfinkle}}\ \emph {et~al.}(1991)\citenamefont
  {{Garfinkle}}, \citenamefont {{Horowitz}},\ and\ \citenamefont
  {{Strominger}}}]{Garfinkle91}%
  \BibitemOpen
  \bibfield  {author} {\bibinfo {author} {\bibfnamefont {D.}~\bibnamefont
  {{Garfinkle}}}, \bibinfo {author} {\bibfnamefont {G.~T.}\ \bibnamefont
  {{Horowitz}}}, \ and\ \bibinfo {author} {\bibfnamefont {A.}~\bibnamefont
  {{Strominger}}},\ }\href {\doibase 10.1103/PhysRevD.43.3140} {\bibfield
  {journal} {\bibinfo  {journal} {Phys. Rev. D}\ }\textbf {\bibinfo {volume}
  {43}},\ \bibinfo {pages} {3140} (\bibinfo {year} {1991})}\BibitemShut
  {NoStop}%
\bibitem [{\citenamefont {{Brill}}\ and\ \citenamefont
  {{Horowitz}}(1991)}]{Brill91}%
  \BibitemOpen
  \bibfield  {author} {\bibinfo {author} {\bibfnamefont {D.}~\bibnamefont
  {{Brill}}}\ and\ \bibinfo {author} {\bibfnamefont {G.~T.}\ \bibnamefont
  {{Horowitz}}},\ }\href {\doibase 10.1016/0370-2693(91)90618-Z} {\bibfield
  {journal} {\bibinfo  {journal} {Physics Letters B}\ }\textbf {\bibinfo
  {volume} {262}},\ \bibinfo {pages} {437} (\bibinfo {year}
  {1991})}\BibitemShut {NoStop}%
\bibitem [{\citenamefont {{Gregory}}\ and\ \citenamefont
  {{Harvey}}(1993)}]{Gregory93a}%
  \BibitemOpen
  \bibfield  {author} {\bibinfo {author} {\bibfnamefont {R.}~\bibnamefont
  {{Gregory}}}\ and\ \bibinfo {author} {\bibfnamefont {J.~A.}\ \bibnamefont
  {{Harvey}}},\ }\href {\doibase 10.1103/PhysRevD.47.2411} {\bibfield
  {journal} {\bibinfo  {journal} {Phys. Rev. D}\ }\textbf {\bibinfo {volume}
  {47}},\ \bibinfo {pages} {2411} (\bibinfo {year} {1993})},\ \Eprint
  {http://arxiv.org/abs/hep-th/9209070} {arXiv:hep-th/9209070 [hep-th]}
  \BibitemShut {NoStop}%
\bibitem [{\citenamefont {{Koikawa}}\ and\ \citenamefont
  {{Yoshimura}}(1987)}]{Koikawa87}%
  \BibitemOpen
  \bibfield  {author} {\bibinfo {author} {\bibfnamefont {T.}~\bibnamefont
  {{Koikawa}}}\ and\ \bibinfo {author} {\bibfnamefont {M.}~\bibnamefont
  {{Yoshimura}}},\ }\href {\doibase 10.1016/0370-2693(87)91264-0} {\bibfield
  {journal} {\bibinfo  {journal} {Physics Letters B}\ }\textbf {\bibinfo
  {volume} {189}},\ \bibinfo {pages} {29} (\bibinfo {year} {1987})}\BibitemShut
  {NoStop}%
\bibitem [{\citenamefont {{Boulware}}\ and\ \citenamefont
  {{Deser}}(1986)}]{Boulware86}%
  \BibitemOpen
  \bibfield  {author} {\bibinfo {author} {\bibfnamefont {D.~G.}\ \bibnamefont
  {{Boulware}}}\ and\ \bibinfo {author} {\bibfnamefont {S.}~\bibnamefont
  {{Deser}}},\ }\href {\doibase 10.1016/0370-2693(86)90614-3} {\bibfield
  {journal} {\bibinfo  {journal} {Physics Letters B}\ }\textbf {\bibinfo
  {volume} {175}},\ \bibinfo {pages} {409} (\bibinfo {year}
  {1986})}\BibitemShut {NoStop}%
\bibitem [{\citenamefont {{Rakhmanov}}(1994)}]{Rakhmanov94}%
  \BibitemOpen
  \bibfield  {author} {\bibinfo {author} {\bibfnamefont {M.}~\bibnamefont
  {{Rakhmanov}}},\ }\href {\doibase 10.1103/PhysRevD.50.5155} {\bibfield
  {journal} {\bibinfo  {journal} {Phys. Rev. D}\ }\textbf {\bibinfo {volume}
  {50}},\ \bibinfo {pages} {5155} (\bibinfo {year} {1994})},\ \Eprint
  {http://arxiv.org/abs/hep-th/9310174} {arXiv:hep-th/9310174 [hep-th]}
  \BibitemShut {NoStop}%
\bibitem [{\citenamefont {{Harms}}\ and\ \citenamefont
  {{Leblanc}}(1992)}]{Harms92}%
  \BibitemOpen
  \bibfield  {author} {\bibinfo {author} {\bibfnamefont {B.}~\bibnamefont
  {{Harms}}}\ and\ \bibinfo {author} {\bibfnamefont {Y.}~\bibnamefont
  {{Leblanc}}},\ }\href {\doibase 10.1103/PhysRevD.46.2334} {\bibfield
  {journal} {\bibinfo  {journal} {Phys. Rev. D}\ }\textbf {\bibinfo {volume}
  {46}},\ \bibinfo {pages} {2334} (\bibinfo {year} {1992})},\ \Eprint
  {http://arxiv.org/abs/hep-th/9205021} {arXiv:hep-th/9205021 [hep-th]}
  \BibitemShut {NoStop}%
\bibitem [{\citenamefont {{Holzhey}}\ and\ \citenamefont
  {{Wilczek}}(1992)}]{Holzhey92}%
  \BibitemOpen
  \bibfield  {author} {\bibinfo {author} {\bibfnamefont {C.~F.~E.}\
  \bibnamefont {{Holzhey}}}\ and\ \bibinfo {author} {\bibfnamefont
  {F.}~\bibnamefont {{Wilczek}}},\ }\href {\doibase
  10.1016/0550-3213(92)90254-9} {\bibfield  {journal} {\bibinfo  {journal}
  {Nuclear Physics B}\ }\textbf {\bibinfo {volume} {380}},\ \bibinfo {pages}
  {447} (\bibinfo {year} {1992})},\ \Eprint
  {http://arxiv.org/abs/hep-th/9202014} {arXiv:hep-th/9202014 [hep-th]}
  \BibitemShut {NoStop}%
\bibitem [{\citenamefont {{Maldacena}}(1998)}]{Maldacena98}%
  \BibitemOpen
  \bibfield  {author} {\bibinfo {author} {\bibfnamefont {J.~M.}\ \bibnamefont
  {{Maldacena}}},\ }\href@noop {} {\bibfield  {journal} {\bibinfo  {journal}
  {Advances in Theoretical and Mathematical Physics}\ }\textbf {\bibinfo
  {volume} {2}},\ \bibinfo {pages} {231} (\bibinfo {year} {1998})}\BibitemShut
  {NoStop}%
\bibitem [{\citenamefont {{Maldacena}}(1999)}]{Maldacena99}%
  \BibitemOpen
  \bibfield  {author} {\bibinfo {author} {\bibfnamefont {J.}~\bibnamefont
  {{Maldacena}}},\ }\href {\doibase 10.1023/A:1026654312961} {\bibfield
  {journal} {\bibinfo  {journal} {International Journal of Theoretical
  Physics}\ }\textbf {\bibinfo {volume} {38}},\ \bibinfo {pages} {1113}
  (\bibinfo {year} {1999})},\ \Eprint {http://arxiv.org/abs/hep-th/9711200}
  {arXiv:hep-th/9711200 [hep-th]} \BibitemShut {NoStop}%
\bibitem [{\citenamefont {{Witten}}(1998)}]{Witten98}%
  \BibitemOpen
  \bibfield  {author} {\bibinfo {author} {\bibfnamefont {E.}~\bibnamefont
  {{Witten}}},\ }\href@noop {} {\bibfield  {journal} {\bibinfo  {journal}
  {Advances in Theoretical and Mathematical Physics}\ }\textbf {\bibinfo
  {volume} {2}},\ \bibinfo {pages} {253} (\bibinfo {year} {1998})},\ \Eprint
  {http://arxiv.org/abs/hep-th/9802150} {arXiv:hep-th/9802150 [hep-th]}
  \BibitemShut {NoStop}%
\bibitem [{\citenamefont {{Klemm}}\ and\ \citenamefont
  {{Sabra}}(2001)}]{Klemm01}%
  \BibitemOpen
  \bibfield  {author} {\bibinfo {author} {\bibfnamefont {D.}~\bibnamefont
  {{Klemm}}}\ and\ \bibinfo {author} {\bibfnamefont {W.~A.}\ \bibnamefont
  {{Sabra}}},\ }\href {\doibase 10.1016/S0370-2693(01)00181-2} {\bibfield
  {journal} {\bibinfo  {journal} {Physics Letters B}\ }\textbf {\bibinfo
  {volume} {503}},\ \bibinfo {pages} {147} (\bibinfo {year} {2001})},\ \Eprint
  {http://arxiv.org/abs/hep-th/0010200} {arXiv:hep-th/0010200 [hep-th]}
  \BibitemShut {NoStop}%
\bibitem [{\citenamefont {{Gubser}}\ \emph {et~al.}(1998)\citenamefont
  {{Gubser}}, \citenamefont {{Klebanov}},\ and\ \citenamefont
  {{Polyakov}}}]{Gubser98}%
  \BibitemOpen
  \bibfield  {author} {\bibinfo {author} {\bibfnamefont {S.~S.}\ \bibnamefont
  {{Gubser}}}, \bibinfo {author} {\bibfnamefont {I.~R.}\ \bibnamefont
  {{Klebanov}}}, \ and\ \bibinfo {author} {\bibfnamefont {A.~M.}\ \bibnamefont
  {{Polyakov}}},\ }\href {\doibase 10.1016/S0370-2693(98)00377-3} {\bibfield
  {journal} {\bibinfo  {journal} {Physics Letters B}\ }\textbf {\bibinfo
  {volume} {428}},\ \bibinfo {pages} {105} (\bibinfo {year} {1998})},\ \Eprint
  {http://arxiv.org/abs/hep-th/9802109} {arXiv:hep-th/9802109 [hep-th]}
  \BibitemShut {NoStop}%
\bibitem [{\citenamefont {{Aharony}}\ \emph {et~al.}(2000)\citenamefont
  {{Aharony}}, \citenamefont {{Gubser}}, \citenamefont {{Maldacena}},
  \citenamefont {{Ooguri}},\ and\ \citenamefont {{Oz}}}]{Aharony00}%
  \BibitemOpen
  \bibfield  {author} {\bibinfo {author} {\bibfnamefont {O.}~\bibnamefont
  {{Aharony}}}, \bibinfo {author} {\bibfnamefont {S.~S.}\ \bibnamefont
  {{Gubser}}}, \bibinfo {author} {\bibfnamefont {J.}~\bibnamefont
  {{Maldacena}}}, \bibinfo {author} {\bibfnamefont {H.}~\bibnamefont
  {{Ooguri}}}, \ and\ \bibinfo {author} {\bibfnamefont {Y.}~\bibnamefont
  {{Oz}}},\ }\href {\doibase 10.1016/S0370-1573(99)00083-6} {\bibfield
  {journal} {\bibinfo  {journal} {Physics Report}\ }\textbf {\bibinfo {volume}
  {323}},\ \bibinfo {pages} {183} (\bibinfo {year} {2000})},\ \Eprint
  {http://arxiv.org/abs/hep-th/9905111} {arXiv:hep-th/9905111 [hep-th]}
  \BibitemShut {NoStop}%
\bibitem [{\citenamefont {Jai-akson}\ \emph {et~al.}(2017)\citenamefont
  {Jai-akson}, \citenamefont {Chatrabhuti}, \citenamefont {Evnin},\ and\
  \citenamefont {Lehner}}]{Jai-akson2017}%
  \BibitemOpen
  \bibfield  {author} {\bibinfo {author} {\bibfnamefont {P.}~\bibnamefont
  {Jai-akson}}, \bibinfo {author} {\bibfnamefont {A.}~\bibnamefont
  {Chatrabhuti}}, \bibinfo {author} {\bibfnamefont {O.}~\bibnamefont {Evnin}},
  \ and\ \bibinfo {author} {\bibfnamefont {L.}~\bibnamefont {Lehner}},\ }\href
  {\doibase 10.1103/PhysRevD.96.044031} {\bibfield  {journal} {\bibinfo
  {journal} {Phys. Rev. D}\ }\textbf {\bibinfo {volume} {96}},\ \bibinfo
  {pages} {044031} (\bibinfo {year} {2017})}\BibitemShut {NoStop}%
\bibitem [{\citenamefont {{Heydari-Fard}}\ \emph {et~al.}(2020)\citenamefont
  {{Heydari-Fard}}, \citenamefont {{Heydari-Fard}},\ and\ \citenamefont
  {{Sepangi}}}]{Heydari-Fard2020}%
  \BibitemOpen
  \bibfield  {author} {\bibinfo {author} {\bibfnamefont {M.}~\bibnamefont
  {{Heydari-Fard}}}, \bibinfo {author} {\bibfnamefont {M.}~\bibnamefont
  {{Heydari-Fard}}}, \ and\ \bibinfo {author} {\bibfnamefont {H.~R.}\
  \bibnamefont {{Sepangi}}},\ }\href {\doibase 10.1140/epjc/s10052-020-7911-0}
  {\bibfield  {journal} {\bibinfo  {journal} {European Physical Journal C}\
  }\textbf {\bibinfo {volume} {80}},\ \bibinfo {eid} {351} (\bibinfo {year}
  {2020})},\ \Eprint {http://arxiv.org/abs/2004.05552} {arXiv:2004.05552
  [gr-qc]} \BibitemShut {NoStop}%
\bibitem [{\citenamefont {{Eubanks}}\ \emph {et~al.}(1997)\citenamefont
  {{Eubanks}}, \citenamefont {{Matsakis}}, \citenamefont {{Martin}},
  \citenamefont {{Archinal}}, \citenamefont {{McCarthy}}, \citenamefont
  {{Klioner}}, \citenamefont {{Shapiro}},\ and\ \citenamefont
  {{Shapiro}}}]{Eubanks97}%
  \BibitemOpen
  \bibfield  {author} {\bibinfo {author} {\bibfnamefont {T.~M.}\ \bibnamefont
  {{Eubanks}}}, \bibinfo {author} {\bibfnamefont {D.~N.}\ \bibnamefont
  {{Matsakis}}}, \bibinfo {author} {\bibfnamefont {J.~O.}\ \bibnamefont
  {{Martin}}}, \bibinfo {author} {\bibfnamefont {B.~A.}\ \bibnamefont
  {{Archinal}}}, \bibinfo {author} {\bibfnamefont {D.~D.}\ \bibnamefont
  {{McCarthy}}}, \bibinfo {author} {\bibfnamefont {S.~A.}\ \bibnamefont
  {{Klioner}}}, \bibinfo {author} {\bibfnamefont {S.}~\bibnamefont
  {{Shapiro}}}, \ and\ \bibinfo {author} {\bibfnamefont {I.~I.}\ \bibnamefont
  {{Shapiro}}},\ }in\ \href@noop {} {\emph {\bibinfo {booktitle} {APS April
  Meeting Abstracts}}}\ (\bibinfo {year} {1997})\BibitemShut {NoStop}%
\bibitem [{\citenamefont {{Iorio}}\ and\ \citenamefont
  {{Ruggiero}}(2010)}]{Iorio10}%
  \BibitemOpen
  \bibfield  {author} {\bibinfo {author} {\bibfnamefont {L.}~\bibnamefont
  {{Iorio}}}\ and\ \bibinfo {author} {\bibfnamefont {M.~L.}\ \bibnamefont
  {{Ruggiero}}},\ }\href {\doibase 10.1142/S0217751X10050780} {\bibfield
  {journal} {\bibinfo  {journal} {International Journal of Modern Physics A}\
  }\textbf {\bibinfo {volume} {25}},\ \bibinfo {pages} {5399} (\bibinfo {year}
  {2010})},\ \Eprint {http://arxiv.org/abs/0909.2562} {arXiv:0909.2562 [gr-qc]}
  \BibitemShut {NoStop}%
\bibitem [{\citenamefont {{Lobo}}\ \emph {et~al.}(2010)\citenamefont {{Lobo}},
  \citenamefont {{Harko}},\ and\ \citenamefont {{Kov{\'a}cs}}}]{Lobo10}%
  \BibitemOpen
  \bibfield  {author} {\bibinfo {author} {\bibfnamefont {F.~S.~N.}\
  \bibnamefont {{Lobo}}}, \bibinfo {author} {\bibfnamefont {T.}~\bibnamefont
  {{Harko}}}, \ and\ \bibinfo {author} {\bibfnamefont {Z.}~\bibnamefont
  {{Kov{\'a}cs}}},\ }\href@noop {} {\bibfield  {journal} {\bibinfo  {journal}
  {ArXiv e-prints}\ } (\bibinfo {year} {2010})},\ \Eprint
  {http://arxiv.org/abs/1001.3517} {arXiv:1001.3517 [gr-qc]} \BibitemShut
  {NoStop}%
\bibitem [{\citenamefont {{Olmo}}(2011)}]{Olmo11}%
  \BibitemOpen
  \bibfield  {author} {\bibinfo {author} {\bibfnamefont {G.~J.}\ \bibnamefont
  {{Olmo}}},\ }\href {\doibase 10.1142/S0218271811018925} {\bibfield  {journal}
  {\bibinfo  {journal} {International Journal of Modern Physics D}\ }\textbf
  {\bibinfo {volume} {20}},\ \bibinfo {pages} {413} (\bibinfo {year} {2011})},\
  \Eprint {http://arxiv.org/abs/1101.3864} {arXiv:1101.3864 [gr-qc]}
  \BibitemShut {NoStop}%
\bibitem [{\citenamefont {{Abbott}}\ \emph {et~al.}(2016)\citenamefont
  {{Abbott}}, \citenamefont {{Abbott}}, \citenamefont {{Abbott}}, \citenamefont
  {{Abernathy}}, \citenamefont {{Acernese}}, \citenamefont {{Ackley}},
  \citenamefont {{Adams}}, \citenamefont {{Adams}}, \citenamefont {{Addesso}},
  \citenamefont {{Adhikari}},\ and\ \citenamefont {et~al.}}]{LIGO16}%
  \BibitemOpen
  \bibfield  {author} {\bibinfo {author} {\bibfnamefont {B.~P.}\ \bibnamefont
  {{Abbott}}}, \bibinfo {author} {\bibfnamefont {R.}~\bibnamefont {{Abbott}}},
  \bibinfo {author} {\bibfnamefont {T.~D.}\ \bibnamefont {{Abbott}}}, \bibinfo
  {author} {\bibfnamefont {M.~R.}\ \bibnamefont {{Abernathy}}}, \bibinfo
  {author} {\bibfnamefont {F.}~\bibnamefont {{Acernese}}}, \bibinfo {author}
  {\bibfnamefont {K.}~\bibnamefont {{Ackley}}}, \bibinfo {author}
  {\bibfnamefont {C.}~\bibnamefont {{Adams}}}, \bibinfo {author} {\bibfnamefont
  {T.}~\bibnamefont {{Adams}}}, \bibinfo {author} {\bibfnamefont
  {P.}~\bibnamefont {{Addesso}}}, \bibinfo {author} {\bibfnamefont {R.~X.}\
  \bibnamefont {{Adhikari}}}, \ and\ \bibinfo {author} {\bibnamefont
  {et~al.}},\ }\href {\doibase 10.1103/PhysRevLett.116.061102} {\bibfield
  {journal} {\bibinfo  {journal} {Physical Review Letters}\ }\textbf {\bibinfo
  {volume} {116}},\ \bibinfo {eid} {061102} (\bibinfo {year} {2016})},\ \Eprint
  {http://arxiv.org/abs/1602.03837} {arXiv:1602.03837 [gr-qc]} \BibitemShut
  {NoStop}%
\bibitem [{\citenamefont {Akiyama}\ \emph
  {et~al.}(2019{\natexlab{a}})\citenamefont {Akiyama} \emph {et~al.}}]{EHT19a}%
  \BibitemOpen
  \bibfield  {author} {\bibinfo {author} {\bibfnamefont {K.}~\bibnamefont
  {Akiyama}} \emph {et~al.} (\bibinfo {collaboration} {Event Horizon
  Telescope}),\ }\href {\doibase 10.3847/2041-8213/ab0ec7} {\bibfield
  {journal} {\bibinfo  {journal} {Astrophys. J.}\ }\textbf {\bibinfo {volume}
  {875}},\ \bibinfo {pages} {L1} (\bibinfo {year}
  {2019}{\natexlab{a}})}\BibitemShut {NoStop}%
%%CITATION = ASJOA,875,L1;%%
\bibitem [{\citenamefont {Akiyama}\ \emph
  {et~al.}(2019{\natexlab{b}})\citenamefont {Akiyama} \emph {et~al.}}]{EHT19b}%
  \BibitemOpen
  \bibfield  {author} {\bibinfo {author} {\bibfnamefont {K.}~\bibnamefont
  {Akiyama}} \emph {et~al.} (\bibinfo {collaboration} {Event Horizon
  Telescope}),\ }\href {\doibase 10.3847/2041-8213/ab1141} {\bibfield
  {journal} {\bibinfo  {journal} {Astrophys. J.}\ }\textbf {\bibinfo {volume}
  {875}},\ \bibinfo {pages} {L6} (\bibinfo {year}
  {2019}{\natexlab{b}})}\BibitemShut {NoStop}%
%%CITATION = ASJOA,875,L6;%%
\bibitem [{\citenamefont {{Bambi}}\ \emph {et~al.}(2016)\citenamefont
  {{Bambi}}, \citenamefont {{Malafarina}},\ and\ \citenamefont
  {{Modesto}}}]{Bambi16c}%
  \BibitemOpen
  \bibfield  {author} {\bibinfo {author} {\bibfnamefont {C.}~\bibnamefont
  {{Bambi}}}, \bibinfo {author} {\bibfnamefont {D.}~\bibnamefont
  {{Malafarina}}}, \ and\ \bibinfo {author} {\bibfnamefont {L.}~\bibnamefont
  {{Modesto}}},\ }\href {\doibase 10.1007/JHEP04(2016)147} {\bibfield
  {journal} {\bibinfo  {journal} {Journal of High Energy Physics}\ }\textbf
  {\bibinfo {volume} {2016}},\ \bibinfo {eid} {147} (\bibinfo {year} {2016})},\
  \Eprint {http://arxiv.org/abs/1603.09592} {arXiv:1603.09592 [gr-qc]}
  \BibitemShut {NoStop}%
\bibitem [{\citenamefont {{Zhou}}\ \emph {et~al.}(2018)\citenamefont {{Zhou}},
  \citenamefont {{Cao}}, \citenamefont {{Abdikamalov}}, \citenamefont
  {{Ayzenberg}}, \citenamefont {{Bambi}}, \citenamefont {{Modesto}},\ and\
  \citenamefont {{Nampalliwar}}}]{Zhou18}%
  \BibitemOpen
  \bibfield  {author} {\bibinfo {author} {\bibfnamefont {M.}~\bibnamefont
  {{Zhou}}}, \bibinfo {author} {\bibfnamefont {Z.}~\bibnamefont {{Cao}}},
  \bibinfo {author} {\bibfnamefont {A.}~\bibnamefont {{Abdikamalov}}}, \bibinfo
  {author} {\bibfnamefont {D.}~\bibnamefont {{Ayzenberg}}}, \bibinfo {author}
  {\bibfnamefont {C.}~\bibnamefont {{Bambi}}}, \bibinfo {author} {\bibfnamefont
  {L.}~\bibnamefont {{Modesto}}}, \ and\ \bibinfo {author} {\bibfnamefont
  {S.}~\bibnamefont {{Nampalliwar}}},\ }\href {\doibase
  10.1103/PhysRevD.98.024007} {\bibfield  {journal} {\bibinfo  {journal} {Phys.
  Rev. D}\ }\textbf {\bibinfo {volume} {98}},\ \bibinfo {eid} {024007}
  (\bibinfo {year} {2018})},\ \Eprint {http://arxiv.org/abs/1803.07849}
  {arXiv:1803.07849 [gr-qc]} \BibitemShut {NoStop}%
\bibitem [{\citenamefont {{Tripathi}}\ \emph {et~al.}(2019)\citenamefont
  {{Tripathi}}, \citenamefont {{Yan}}, \citenamefont {{Yang}}, \citenamefont
  {{Yan}}, \citenamefont {{Garnham}}, \citenamefont {{Yao}}, \citenamefont
  {{Li}}, \citenamefont {{Ding}}, \citenamefont {{Abdikamalov}}, \citenamefont
  {{Ayzenberg}}, \citenamefont {{Bambi}}, \citenamefont {{Dauser}},
  \citenamefont {{Garcia}}, \citenamefont {{Jiang}},\ and\ \citenamefont
  {{Nampalliwar}}}]{Tripathi19}%
  \BibitemOpen
  \bibfield  {author} {\bibinfo {author} {\bibfnamefont {A.}~\bibnamefont
  {{Tripathi}}}, \bibinfo {author} {\bibfnamefont {J.}~\bibnamefont {{Yan}}},
  \bibinfo {author} {\bibfnamefont {Y.}~\bibnamefont {{Yang}}}, \bibinfo
  {author} {\bibfnamefont {Y.}~\bibnamefont {{Yan}}}, \bibinfo {author}
  {\bibfnamefont {M.}~\bibnamefont {{Garnham}}}, \bibinfo {author}
  {\bibfnamefont {Y.}~\bibnamefont {{Yao}}}, \bibinfo {author} {\bibfnamefont
  {S.}~\bibnamefont {{Li}}}, \bibinfo {author} {\bibfnamefont {Z.}~\bibnamefont
  {{Ding}}}, \bibinfo {author} {\bibfnamefont {A.~B.}\ \bibnamefont
  {{Abdikamalov}}}, \bibinfo {author} {\bibfnamefont {D.}~\bibnamefont
  {{Ayzenberg}}}, \bibinfo {author} {\bibfnamefont {C.}~\bibnamefont
  {{Bambi}}}, \bibinfo {author} {\bibfnamefont {T.}~\bibnamefont {{Dauser}}},
  \bibinfo {author} {\bibfnamefont {J.~A.}\ \bibnamefont {{Garcia}}}, \bibinfo
  {author} {\bibfnamefont {J.}~\bibnamefont {{Jiang}}}, \ and\ \bibinfo
  {author} {\bibfnamefont {S.}~\bibnamefont {{Nampalliwar}}},\ }\href@noop {}
  {\bibfield  {journal} {\bibinfo  {journal} {arXiv e-prints}\ } (\bibinfo
  {year} {2019})},\ \Eprint {http://arxiv.org/abs/1901.03064} {arXiv:1901.03064
  [gr-qc]} \BibitemShut {NoStop}%
\bibitem [{\citenamefont {{Abramowicz}}\ and\ \citenamefont
  {{Klu{\'z}niak}}(2001)}]{Abramowicz2001}%
  \BibitemOpen
  \bibfield  {author} {\bibinfo {author} {\bibfnamefont {M.~A.}\ \bibnamefont
  {{Abramowicz}}}\ and\ \bibinfo {author} {\bibfnamefont {W.}~\bibnamefont
  {{Klu{\'z}niak}}},\ }\href {\doibase 10.1051/0004-6361:20010791} {\bibfield
  {journal} {\bibinfo  {journal} {Astron. Astrophys.}\ }\textbf {\bibinfo
  {volume} {374}},\ \bibinfo {pages} {L19} (\bibinfo {year} {2001})},\ \Eprint
  {http://arxiv.org/abs/arXiv:astro-ph/0105077} {arXiv:astro-ph/0105077}
  \BibitemShut {NoStop}%
\bibitem [{\citenamefont {{T{\"o}r{\"o}k}}(2005)}]{Torok2005}%
  \BibitemOpen
  \bibfield  {author} {\bibinfo {author} {\bibfnamefont {G.}~\bibnamefont
  {{T{\"o}r{\"o}k}}},\ }\href {\doibase 10.1051/0004-6361:20042558} {\bibfield
  {journal} {\bibinfo  {journal} {Astron. Astrophys.}\ }\textbf {\bibinfo
  {volume} {440}},\ \bibinfo {pages} {1} (\bibinfo {year} {2005})},\ \Eprint
  {http://arxiv.org/abs/arXiv:astro-ph/0412500} {arXiv:astro-ph/0412500}
  \BibitemShut {NoStop}%
\bibitem [{\citenamefont {{T{\"o}r{\"o}k}}\ \emph {et~al.}(2005)\citenamefont
  {{T{\"o}r{\"o}k}}, \citenamefont {{Abramowicz}}, \citenamefont
  {{Klu{\'z}niak}},\ and\ \citenamefont {{Stuchl{\'{\i}}k}}}]{Torok2005b}%
  \BibitemOpen
  \bibfield  {author} {\bibinfo {author} {\bibfnamefont {G.}~\bibnamefont
  {{T{\"o}r{\"o}k}}}, \bibinfo {author} {\bibfnamefont {M.~A.}\ \bibnamefont
  {{Abramowicz}}}, \bibinfo {author} {\bibfnamefont {W.}~\bibnamefont
  {{Klu{\'z}niak}}}, \ and\ \bibinfo {author} {\bibfnamefont {Z.}~\bibnamefont
  {{Stuchl{\'{\i}}k}}},\ }\href {\doibase 10.1051/0004-6361:20047115}
  {\bibfield  {journal} {\bibinfo  {journal} {Astron. Astrophys.}\ }\textbf
  {\bibinfo {volume} {436}},\ \bibinfo {pages} {1} (\bibinfo {year}
  {2005})}\BibitemShut {NoStop}%
\bibitem [{\citenamefont {{Stuchl{\'\i}k}}\ \emph {et~al.}(2013)\citenamefont
  {{Stuchl{\'\i}k}}, \citenamefont {{Kotrlov{\'a}}},\ and\ \citenamefont
  {{T{\"o}r{\"o}k}}}]{Stuchlik2013}%
  \BibitemOpen
  \bibfield  {author} {\bibinfo {author} {\bibfnamefont {Z.}~\bibnamefont
  {{Stuchl{\'\i}k}}}, \bibinfo {author} {\bibfnamefont {A.}~\bibnamefont
  {{Kotrlov{\'a}}}}, \ and\ \bibinfo {author} {\bibfnamefont {G.}~\bibnamefont
  {{T{\"o}r{\"o}k}}},\ }\href {\doibase 10.1051/0004-6361/201219724} {\bibfield
   {journal} {\bibinfo  {journal} {Astrn. Astrophys.}\ }\textbf {\bibinfo
  {volume} {552}},\ \bibinfo {eid} {A10} (\bibinfo {year} {2013})},\ \Eprint
  {http://arxiv.org/abs/1305.3552} {arXiv:1305.3552 [astro-ph.HE]} \BibitemShut
  {NoStop}%
\bibitem [{\citenamefont {{Tursunov}}\ \emph {et~al.}(2016)\citenamefont
  {{Tursunov}}, \citenamefont {{Stuchl{\'{\i}}k}},\ and\ \citenamefont
  {{Kolo{\v s}}}}]{Tursunov16}%
  \BibitemOpen
  \bibfield  {author} {\bibinfo {author} {\bibfnamefont {A.}~\bibnamefont
  {{Tursunov}}}, \bibinfo {author} {\bibfnamefont {Z.}~\bibnamefont
  {{Stuchl{\'{\i}}k}}}, \ and\ \bibinfo {author} {\bibfnamefont
  {M.}~\bibnamefont {{Kolo{\v s}}}},\ }\href {\doibase
  10.1103/PhysRevD.93.084012} {\bibfield  {journal} {\bibinfo  {journal} {Phys.
  Rev. D}\ }\textbf {\bibinfo {volume} {93}},\ \bibinfo {eid} {084012}
  (\bibinfo {year} {2016})},\ \Eprint {http://arxiv.org/abs/1603.07264}
  {arXiv:1603.07264 [gr-qc]} \BibitemShut {NoStop}%
\bibitem [{\citenamefont {{Kolo{\v s}}}\ \emph {et~al.}(2017)\citenamefont
  {{Kolo{\v s}}}, \citenamefont {{Tursunov}},\ and\ \citenamefont
  {{Stuchl{\'{\i}}k}}}]{Kolos17}%
  \BibitemOpen
  \bibfield  {author} {\bibinfo {author} {\bibfnamefont {M.}~\bibnamefont
  {{Kolo{\v s}}}}, \bibinfo {author} {\bibfnamefont {A.}~\bibnamefont
  {{Tursunov}}}, \ and\ \bibinfo {author} {\bibfnamefont {Z.}~\bibnamefont
  {{Stuchl{\'{\i}}k}}},\ }\href@noop {} {\bibfield  {journal} {\bibinfo
  {journal} {Eur. Phys. J. C.}\ }\textbf {\bibinfo {volume} {77}},\ \bibinfo
  {pages} {860} (\bibinfo {year} {2017})},\ \Eprint
  {http://arxiv.org/abs/1707.02224} {arXiv:1707.02224 [astro-ph.HE]}
  \BibitemShut {NoStop}%
\bibitem [{\citenamefont {{Steiner}}\ \emph {et~al.}(2011)\citenamefont
  {{Steiner}}, \citenamefont {{Reis}}, \citenamefont {{McClintock}},
  \citenamefont {{Narayan}}, \citenamefont {{Remillard}}, \citenamefont
  {{Orosz}}, \citenamefont {{Gou}}, \citenamefont {{Fabian}},\ and\
  \citenamefont {{Torres}}}]{Steiner11}%
  \BibitemOpen
  \bibfield  {author} {\bibinfo {author} {\bibfnamefont {J.~F.}\ \bibnamefont
  {{Steiner}}}, \bibinfo {author} {\bibfnamefont {R.~C.}\ \bibnamefont
  {{Reis}}}, \bibinfo {author} {\bibfnamefont {J.~E.}\ \bibnamefont
  {{McClintock}}}, \bibinfo {author} {\bibfnamefont {R.}~\bibnamefont
  {{Narayan}}}, \bibinfo {author} {\bibfnamefont {R.~A.}\ \bibnamefont
  {{Remillard}}}, \bibinfo {author} {\bibfnamefont {J.~A.}\ \bibnamefont
  {{Orosz}}}, \bibinfo {author} {\bibfnamefont {L.}~\bibnamefont {{Gou}}},
  \bibinfo {author} {\bibfnamefont {A.~C.}\ \bibnamefont {{Fabian}}}, \ and\
  \bibinfo {author} {\bibfnamefont {M.~A.~P.}\ \bibnamefont {{Torres}}},\
  }\href {\doibase 10.1111/j.1365-2966.2011.19089.x} {\bibfield  {journal}
  {\bibinfo  {journal} {Mon. Not. R. Astron. Soc}\ }\textbf {\bibinfo {volume}
  {416}},\ \bibinfo {pages} {941} (\bibinfo {year} {2011})},\ \Eprint
  {http://arxiv.org/abs/1010.1013} {arXiv:1010.1013 [astro-ph.HE]} \BibitemShut
  {NoStop}%
\bibitem [{\citenamefont {{Gou}}\ \emph {et~al.}(2014)\citenamefont {{Gou}},
  \citenamefont {{McClintock}}, \citenamefont {{Remillard}}, \citenamefont
  {{Steiner}}, \citenamefont {{Reid}}, \citenamefont {{Orosz}}, \citenamefont
  {{Narayan}}, \citenamefont {{Hanke}},\ and\ \citenamefont
  {{Garc{\'{\i}}a}}}]{Gou14}%
  \BibitemOpen
  \bibfield  {author} {\bibinfo {author} {\bibfnamefont {L.}~\bibnamefont
  {{Gou}}}, \bibinfo {author} {\bibfnamefont {J.~E.}\ \bibnamefont
  {{McClintock}}}, \bibinfo {author} {\bibfnamefont {R.~A.}\ \bibnamefont
  {{Remillard}}}, \bibinfo {author} {\bibfnamefont {J.~F.}\ \bibnamefont
  {{Steiner}}}, \bibinfo {author} {\bibfnamefont {M.~J.}\ \bibnamefont
  {{Reid}}}, \bibinfo {author} {\bibfnamefont {J.~A.}\ \bibnamefont {{Orosz}}},
  \bibinfo {author} {\bibfnamefont {R.}~\bibnamefont {{Narayan}}}, \bibinfo
  {author} {\bibfnamefont {M.}~\bibnamefont {{Hanke}}}, \ and\ \bibinfo
  {author} {\bibfnamefont {J.}~\bibnamefont {{Garc{\'{\i}}a}}},\ }\href
  {\doibase 10.1088/0004-637X/790/1/29} {\bibfield  {journal} {\bibinfo
  {journal} {Astrophys. J.}\ }\textbf {\bibinfo {volume} {790}},\ \bibinfo
  {eid} {29} (\bibinfo {year} {2014})},\ \Eprint
  {http://arxiv.org/abs/1308.4760} {arXiv:1308.4760 [astro-ph.HE]} \BibitemShut
  {NoStop}%
\bibitem [{\citenamefont {{McClintock}}\ \emph {et~al.}(2014)\citenamefont
  {{McClintock}}, \citenamefont {{Narayan}},\ and\ \citenamefont
  {{Steiner}}}]{McClintock14}%
  \BibitemOpen
  \bibfield  {author} {\bibinfo {author} {\bibfnamefont {J.~E.}\ \bibnamefont
  {{McClintock}}}, \bibinfo {author} {\bibfnamefont {R.}~\bibnamefont
  {{Narayan}}}, \ and\ \bibinfo {author} {\bibfnamefont {J.~F.}\ \bibnamefont
  {{Steiner}}},\ }\href {\doibase 10.1007/s11214-013-0003-9} {\bibfield
  {journal} {\bibinfo  {journal} {Space Science Reviews}\ }\textbf {\bibinfo
  {volume} {183}},\ \bibinfo {pages} {295} (\bibinfo {year} {2014})},\ \Eprint
  {http://arxiv.org/abs/1303.1583} {arXiv:1303.1583 [astro-ph.HE]} \BibitemShut
  {NoStop}%
\bibitem [{\citenamefont {{Steiner}}\ \emph {et~al.}(2010)\citenamefont
  {{Steiner}}, \citenamefont {{McClintock}}, \citenamefont {{Remillard}},
  \citenamefont {{Gou}}, \citenamefont {{Yamada}},\ and\ \citenamefont
  {{Narayan}}}]{Steiner10}%
  \BibitemOpen
  \bibfield  {author} {\bibinfo {author} {\bibfnamefont {J.~F.}\ \bibnamefont
  {{Steiner}}}, \bibinfo {author} {\bibfnamefont {J.~E.}\ \bibnamefont
  {{McClintock}}}, \bibinfo {author} {\bibfnamefont {R.~A.}\ \bibnamefont
  {{Remillard}}}, \bibinfo {author} {\bibfnamefont {L.}~\bibnamefont {{Gou}}},
  \bibinfo {author} {\bibfnamefont {S.}~\bibnamefont {{Yamada}}}, \ and\
  \bibinfo {author} {\bibfnamefont {R.}~\bibnamefont {{Narayan}}},\ }\href
  {\doibase 10.1088/2041-8205/718/2/L117} {\bibfield  {journal} {\bibinfo
  {journal} {Astrophys. J. Lett.}\ }\textbf {\bibinfo {volume} {718}},\
  \bibinfo {pages} {L117} (\bibinfo {year} {2010})},\ \Eprint
  {http://arxiv.org/abs/1006.5729} {arXiv:1006.5729 [astro-ph.HE]} \BibitemShut
  {NoStop}%
\bibitem [{\citenamefont {{Chen}}\ \emph {et~al.}(2016)\citenamefont {{Chen}},
  \citenamefont {{Wang}},\ and\ \citenamefont {{Jing}}}]{Chen16}%
  \BibitemOpen
  \bibfield  {author} {\bibinfo {author} {\bibfnamefont {S.}~\bibnamefont
  {{Chen}}}, \bibinfo {author} {\bibfnamefont {M.}~\bibnamefont {{Wang}}}, \
  and\ \bibinfo {author} {\bibfnamefont {J.}~\bibnamefont {{Jing}}},\ }\href
  {\doibase 10.1007/JHEP09(2016)082} {\bibfield  {journal} {\bibinfo  {journal}
  {Journal of High Energy Physics}\ }\textbf {\bibinfo {volume} {2016}},\
  \bibinfo {eid} {82} (\bibinfo {year} {2016})},\ \Eprint
  {http://arxiv.org/abs/1604.02785} {arXiv:1604.02785 [gr-qc]} \BibitemShut
  {NoStop}%
\bibitem [{\citenamefont {{Hashimoto}}\ and\ \citenamefont
  {{Tanahashi}}(2017)}]{Hashimoto17}%
  \BibitemOpen
  \bibfield  {author} {\bibinfo {author} {\bibfnamefont {K.}~\bibnamefont
  {{Hashimoto}}}\ and\ \bibinfo {author} {\bibfnamefont {N.}~\bibnamefont
  {{Tanahashi}}},\ }\href {\doibase 10.1103/PhysRevD.95.024007} {\bibfield
  {journal} {\bibinfo  {journal} {Phys. rev. D}\ }\textbf {\bibinfo {volume}
  {95}},\ \bibinfo {eid} {024007} (\bibinfo {year} {2017})},\ \Eprint
  {http://arxiv.org/abs/1610.06070} {arXiv:1610.06070 [hep-th]} \BibitemShut
  {NoStop}%
\bibitem [{\citenamefont {{Dalui}}\ \emph {et~al.}(2019)\citenamefont
  {{Dalui}}, \citenamefont {{Majhi}},\ and\ \citenamefont
  {{Mishra}}}]{Dalui19}%
  \BibitemOpen
  \bibfield  {author} {\bibinfo {author} {\bibfnamefont {S.}~\bibnamefont
  {{Dalui}}}, \bibinfo {author} {\bibfnamefont {B.~R.}\ \bibnamefont
  {{Majhi}}}, \ and\ \bibinfo {author} {\bibfnamefont {P.}~\bibnamefont
  {{Mishra}}},\ }\href {\doibase 10.1016/j.physletb.2018.11.050} {\bibfield
  {journal} {\bibinfo  {journal} {Physics Letters B}\ }\textbf {\bibinfo
  {volume} {788}},\ \bibinfo {pages} {486} (\bibinfo {year} {2019})},\ \Eprint
  {http://arxiv.org/abs/1803.06527} {arXiv:1803.06527 [gr-qc]} \BibitemShut
  {NoStop}%
\bibitem [{\citenamefont {{Han}}(2008)}]{Han08}%
  \BibitemOpen
  \bibfield  {author} {\bibinfo {author} {\bibfnamefont {W.}~\bibnamefont
  {{Han}}},\ }\href {\doibase 10.1007/s10714-007-0598-9} {\bibfield  {journal}
  {\bibinfo  {journal} {General Relativity and Gravitation}\ }\textbf {\bibinfo
  {volume} {40}},\ \bibinfo {pages} {1831} (\bibinfo {year} {2008})},\ \Eprint
  {http://arxiv.org/abs/1006.2229} {arXiv:1006.2229 [gr-qc]} \BibitemShut
  {NoStop}%
\bibitem [{\citenamefont {{de Moura}}\ and\ \citenamefont
  {{Letelier}}(2000)}]{Moura00}%
  \BibitemOpen
  \bibfield  {author} {\bibinfo {author} {\bibfnamefont {A.~P.~S.}\
  \bibnamefont {{de Moura}}}\ and\ \bibinfo {author} {\bibfnamefont {P.~S.}\
  \bibnamefont {{Letelier}}},\ }\href {\doibase 10.1103/PhysRevE.61.6506}
  {\bibfield  {journal} {\bibinfo  {journal} {Phys. Rev. E}\ }\textbf {\bibinfo
  {volume} {61}},\ \bibinfo {pages} {6506} (\bibinfo {year} {2000})},\ \Eprint
  {http://arxiv.org/abs/chao-dyn/9910035} {arXiv:chao-dyn/9910035 [nlin.CD]}
  \BibitemShut {NoStop}%
\bibitem [{\citenamefont {{Morozova}}\ \emph {et~al.}(2014)\citenamefont
  {{Morozova}}, \citenamefont {{Rezzolla}},\ and\ \citenamefont
  {{Ahmedov}}}]{Morozova14}%
  \BibitemOpen
  \bibfield  {author} {\bibinfo {author} {\bibfnamefont {V.~S.}\ \bibnamefont
  {{Morozova}}}, \bibinfo {author} {\bibfnamefont {L.}~\bibnamefont
  {{Rezzolla}}}, \ and\ \bibinfo {author} {\bibfnamefont {B.~J.}\ \bibnamefont
  {{Ahmedov}}},\ }\href {\doibase 10.1103/PhysRevD.89.104030} {\bibfield
  {journal} {\bibinfo  {journal} {Phys. Rev. D}\ }\textbf {\bibinfo {volume}
  {89}},\ \bibinfo {eid} {104030} (\bibinfo {year} {2014})},\ \Eprint
  {http://arxiv.org/abs/1310.3575} {arXiv:1310.3575 [gr-qc]} \BibitemShut
  {NoStop}%
\bibitem [{\citenamefont {{Stuchl{\'\i}k}}\ \emph {et~al.}(2020)\citenamefont
  {{Stuchl{\'\i}k}}, \citenamefont {{Kolo{\v{s}}}}, \citenamefont
  {{Kov{\'a}{\v{r}}}}, \citenamefont {{Slan{\'y}}},\ and\ \citenamefont
  {{Tursunov}}}]{Stuchlik2020Univ}%
  \BibitemOpen
  \bibfield  {author} {\bibinfo {author} {\bibfnamefont {Z.}~\bibnamefont
  {{Stuchl{\'\i}k}}}, \bibinfo {author} {\bibfnamefont {M.}~\bibnamefont
  {{Kolo{\v{s}}}}}, \bibinfo {author} {\bibfnamefont {J.}~\bibnamefont
  {{Kov{\'a}{\v{r}}}}}, \bibinfo {author} {\bibfnamefont {P.}~\bibnamefont
  {{Slan{\'y}}}}, \ and\ \bibinfo {author} {\bibfnamefont {A.}~\bibnamefont
  {{Tursunov}}},\ }\href {\doibase 10.3390/universe6020026} {\bibfield
  {journal} {\bibinfo  {journal} {Universe}\ }\textbf {\bibinfo {volume} {6}},\
  \bibinfo {pages} {26} (\bibinfo {year} {2020})}\BibitemShut {NoStop}%
\bibitem [{\citenamefont {{Pugliese}}\ \emph {et~al.}(2010)\citenamefont
  {{Pugliese}}, \citenamefont {{Quevedo}},\ and\ \citenamefont
  {{Ruffini}}}]{Pugliese10}%
  \BibitemOpen
  \bibfield  {author} {\bibinfo {author} {\bibfnamefont {D.}~\bibnamefont
  {{Pugliese}}}, \bibinfo {author} {\bibfnamefont {H.}~\bibnamefont
  {{Quevedo}}}, \ and\ \bibinfo {author} {\bibfnamefont {R.}~\bibnamefont
  {{Ruffini}}},\ }\href@noop {} {\bibfield  {journal} {\bibinfo  {journal}
  {ArXiv e-prints}\ } (\bibinfo {year} {2010})},\ \Eprint
  {http://arxiv.org/abs/1003.2687} {arXiv:1003.2687 [gr-qc]} \BibitemShut
  {NoStop}%
\bibitem [{\citenamefont {{Pugliese}}\ \emph {et~al.}(2011)\citenamefont
  {{Pugliese}}, \citenamefont {{Quevedo}},\ and\ \citenamefont
  {{Ruffini}}}]{Pugliese11}%
  \BibitemOpen
  \bibfield  {author} {\bibinfo {author} {\bibfnamefont {D.}~\bibnamefont
  {{Pugliese}}}, \bibinfo {author} {\bibfnamefont {H.}~\bibnamefont
  {{Quevedo}}}, \ and\ \bibinfo {author} {\bibfnamefont {R.}~\bibnamefont
  {{Ruffini}}},\ }\href {\doibase 10.1103/PhysRevD.83.104052} {\bibfield
  {journal} {\bibinfo  {journal} {Phys. Rev. D}\ }\textbf {\bibinfo {volume}
  {83}},\ \bibinfo {eid} {104052} (\bibinfo {year} {2011})},\ \Eprint
  {http://arxiv.org/abs/1103.1807} {arXiv:1103.1807 [gr-qc]} \BibitemShut
  {NoStop}%
\bibitem [{\citenamefont {{Jawad}}\ \emph {et~al.}(2016)\citenamefont
  {{Jawad}}, \citenamefont {{Ali}}, \citenamefont {{Jamil}},\ and\
  \citenamefont {{Debnath}}}]{Jawad16}%
  \BibitemOpen
  \bibfield  {author} {\bibinfo {author} {\bibfnamefont {A.}~\bibnamefont
  {{Jawad}}}, \bibinfo {author} {\bibfnamefont {F.}~\bibnamefont {{Ali}}},
  \bibinfo {author} {\bibfnamefont {M.}~\bibnamefont {{Jamil}}}, \ and\
  \bibinfo {author} {\bibfnamefont {U.}~\bibnamefont {{Debnath}}},\ }\href
  {\doibase 10.1088/0253-6102/66/5/509} {\bibfield  {journal} {\bibinfo
  {journal} {Communications in Theoretical Physics}\ }\textbf {\bibinfo
  {volume} {66}},\ \bibinfo {pages} {509} (\bibinfo {year} {2016})},\ \Eprint
  {http://arxiv.org/abs/1610.07411} {arXiv:1610.07411 [gr-qc]} \BibitemShut
  {NoStop}%
\bibitem [{\citenamefont {{Hussain}}\ and\ \citenamefont
  {{Jamil}}(2015)}]{Hussain15}%
  \BibitemOpen
  \bibfield  {author} {\bibinfo {author} {\bibfnamefont {S.}~\bibnamefont
  {{Hussain}}}\ and\ \bibinfo {author} {\bibfnamefont {M.}~\bibnamefont
  {{Jamil}}},\ }\href {\doibase 10.1103/PhysRevD.92.043008} {\bibfield
  {journal} {\bibinfo  {journal} {Phys. Rev. D}\ }\textbf {\bibinfo {volume}
  {92}},\ \bibinfo {eid} {043008} (\bibinfo {year} {2015})},\ \Eprint
  {http://arxiv.org/abs/1508.02123} {arXiv:1508.02123 [gr-qc]} \BibitemShut
  {NoStop}%
\bibitem [{\citenamefont {{Jamil}}\ \emph {et~al.}(2015)\citenamefont
  {{Jamil}}, \citenamefont {{Hussain}},\ and\ \citenamefont
  {{Majeed}}}]{Jamil15}%
  \BibitemOpen
  \bibfield  {author} {\bibinfo {author} {\bibfnamefont {M.}~\bibnamefont
  {{Jamil}}}, \bibinfo {author} {\bibfnamefont {S.}~\bibnamefont {{Hussain}}},
  \ and\ \bibinfo {author} {\bibfnamefont {B.}~\bibnamefont {{Majeed}}},\
  }\href {\doibase 10.1140/epjc/s10052-014-3230-7} {\bibfield  {journal}
  {\bibinfo  {journal} {European Physical Journal C}\ }\textbf {\bibinfo
  {volume} {75}},\ \bibinfo {eid} {24} (\bibinfo {year} {2015})},\ \Eprint
  {http://arxiv.org/abs/1404.7123} {arXiv:1404.7123 [gr-qc]} \BibitemShut
  {NoStop}%
\bibitem [{\citenamefont {{Hussain, S}}\ \emph {et~al.}(2014)\citenamefont
  {{Hussain, S}}, \citenamefont {{Hussain, I}},\ and\ \citenamefont {{Jamil,
  M.}}}]{Hussain17}%
  \BibitemOpen
  \bibfield  {author} {\bibinfo {author} {\bibnamefont {{Hussain, S}}},
  \bibinfo {author} {\bibnamefont {{Hussain, I}}}, \ and\ \bibinfo {author}
  {\bibnamefont {{Jamil, M.}}},\ }\href {\doibase
  10.1140/epjc/s10052-014-3210-y} {\bibfield  {journal} {\bibinfo  {journal}
  {Eur. Phys. J. C}\ }\textbf {\bibinfo {volume} {74}},\ \bibinfo {pages} {210}
  (\bibinfo {year} {2014})}\BibitemShut {NoStop}%
\bibitem [{\citenamefont {{Babar}}\ \emph {et~al.}(2016)\citenamefont
  {{Babar}}, \citenamefont {{Jamil}},\ and\ \citenamefont {{Lim}}}]{Babar16}%
  \BibitemOpen
  \bibfield  {author} {\bibinfo {author} {\bibfnamefont {G.~Z.}\ \bibnamefont
  {{Babar}}}, \bibinfo {author} {\bibfnamefont {M.}~\bibnamefont {{Jamil}}}, \
  and\ \bibinfo {author} {\bibfnamefont {Y.-K.}\ \bibnamefont {{Lim}}},\ }\href
  {\doibase 10.1142/S0218271816500243} {\bibfield  {journal} {\bibinfo
  {journal} {International Journal of Modern Physics D}\ }\textbf {\bibinfo
  {volume} {25}},\ \bibinfo {eid} {1650024} (\bibinfo {year} {2016})},\ \Eprint
  {http://arxiv.org/abs/1504.00072} {arXiv:1504.00072 [gr-qc]} \BibitemShut
  {NoStop}%
\bibitem [{\citenamefont {{Ba{\~n}ados}}\ \emph {et~al.}(2009)\citenamefont
  {{Ba{\~n}ados}}, \citenamefont {{Silk}},\ and\ \citenamefont
  {{West}}}]{Banados09}%
  \BibitemOpen
  \bibfield  {author} {\bibinfo {author} {\bibfnamefont {M.}~\bibnamefont
  {{Ba{\~n}ados}}}, \bibinfo {author} {\bibfnamefont {J.}~\bibnamefont
  {{Silk}}}, \ and\ \bibinfo {author} {\bibfnamefont {S.~M.}\ \bibnamefont
  {{West}}},\ }\href {\doibase 10.1103/PhysRevLett.103.111102} {\bibfield
  {journal} {\bibinfo  {journal} {Physical Review Letters}\ }\textbf {\bibinfo
  {volume} {103}},\ \bibinfo {eid} {111102} (\bibinfo {year}
  {2009})}\BibitemShut {NoStop}%
\bibitem [{\citenamefont {{Majeed}}\ and\ \citenamefont
  {{Jamil}}(2017)}]{Majeed17}%
  \BibitemOpen
  \bibfield  {author} {\bibinfo {author} {\bibfnamefont {B.}~\bibnamefont
  {{Majeed}}}\ and\ \bibinfo {author} {\bibfnamefont {M.}~\bibnamefont
  {{Jamil}}},\ }\href {\doibase 10.1142/S0218271817410176} {\bibfield
  {journal} {\bibinfo  {journal} {International Journal of Modern Physics D}\
  }\textbf {\bibinfo {volume} {26}},\ \bibinfo {eid} {1741017} (\bibinfo {year}
  {2017})},\ \Eprint {http://arxiv.org/abs/1705.04167} {arXiv:1705.04167
  [gr-qc]} \BibitemShut {NoStop}%
\bibitem [{\citenamefont {{Zakria}}\ and\ \citenamefont
  {{Jamil}}(2015)}]{Zakria15}%
  \BibitemOpen
  \bibfield  {author} {\bibinfo {author} {\bibfnamefont {A.}~\bibnamefont
  {{Zakria}}}\ and\ \bibinfo {author} {\bibfnamefont {M.}~\bibnamefont
  {{Jamil}}},\ }\href {\doibase 10.1007/JHEP05(2015)147} {\bibfield  {journal}
  {\bibinfo  {journal} {Journal of High Energy Physics}\ }\textbf {\bibinfo
  {volume} {2015}},\ \bibinfo {eid} {147} (\bibinfo {year} {2015})},\ \Eprint
  {http://arxiv.org/abs/1501.06306} {arXiv:1501.06306 [gr-qc]} \BibitemShut
  {NoStop}%
\bibitem [{\citenamefont {{Brevik}}\ and\ \citenamefont
  {{Jamil}}(2019)}]{Brevik19}%
  \BibitemOpen
  \bibfield  {author} {\bibinfo {author} {\bibfnamefont {I.}~\bibnamefont
  {{Brevik}}}\ and\ \bibinfo {author} {\bibfnamefont {M.}~\bibnamefont
  {{Jamil}}},\ }\href {\doibase 10.1142/S0219887819500300} {\bibfield
  {journal} {\bibinfo  {journal} {International Journal of Geometric Methods in
  Modern Physics}\ }\textbf {\bibinfo {volume} {16}},\ \bibinfo {eid} {1950030}
  (\bibinfo {year} {2019})},\ \Eprint {http://arxiv.org/abs/1901.00002}
  {arXiv:1901.00002 [gr-qc]} \BibitemShut {NoStop}%
\bibitem [{\citenamefont {{De Laurentis}}\ \emph {et~al.}(2018)\citenamefont
  {{De Laurentis}}, \citenamefont {{Younsi}}, \citenamefont {{Porth}},
  \citenamefont {{Mizuno}},\ and\ \citenamefont
  {{Rezzolla}}}]{DeLaurentis2018PhRvD}%
  \BibitemOpen
  \bibfield  {author} {\bibinfo {author} {\bibfnamefont {M.}~\bibnamefont {{De
  Laurentis}}}, \bibinfo {author} {\bibfnamefont {Z.}~\bibnamefont {{Younsi}}},
  \bibinfo {author} {\bibfnamefont {O.}~\bibnamefont {{Porth}}}, \bibinfo
  {author} {\bibfnamefont {Y.}~\bibnamefont {{Mizuno}}}, \ and\ \bibinfo
  {author} {\bibfnamefont {L.}~\bibnamefont {{Rezzolla}}},\ }\href {\doibase
  10.1103/PhysRevD.97.104024} {\bibfield  {journal} {\bibinfo  {journal} {Phys.
  Rev. D}\ }\textbf {\bibinfo {volume} {97}},\ \bibinfo {eid} {104024}
  (\bibinfo {year} {2018})},\ \Eprint {http://arxiv.org/abs/1712.00265}
  {arXiv:1712.00265 [gr-qc]} \BibitemShut {NoStop}%
\bibitem [{\citenamefont {Turimov}\ \emph {et~al.}(2018)\citenamefont
  {Turimov}, \citenamefont {Ahmedov}, \citenamefont {Kološ},\ and\
  \citenamefont {Stuchlík}}]{Turimov18a}%
  \BibitemOpen
  \bibfield  {author} {\bibinfo {author} {\bibfnamefont {B.}~\bibnamefont
  {Turimov}}, \bibinfo {author} {\bibfnamefont {B.}~\bibnamefont {Ahmedov}},
  \bibinfo {author} {\bibfnamefont {M.}~\bibnamefont {Kološ}}, \ and\ \bibinfo
  {author} {\bibfnamefont {Z.}~\bibnamefont {Stuchlík}},\ }\href {\doibase
  10.1103/PhysRevD.98.084039} {\bibfield  {journal} {\bibinfo  {journal} {Phys.
  Rev. D}\ }\textbf {\bibinfo {volume} {98}},\ \bibinfo {pages} {084039}
  (\bibinfo {year} {2018})},\ \Eprint {http://arxiv.org/abs/1810.01460}
  {arXiv:1810.01460 [gr-qc]} \BibitemShut {NoStop}%
%%CITATION = ARXIV:1810.01460;%%
\bibitem [{\citenamefont {{de Felice}}\ and\ \citenamefont
  {{Sorge}}(2003)}]{deFelice}%
  \BibitemOpen
  \bibfield  {author} {\bibinfo {author} {\bibfnamefont {F.}~\bibnamefont {{de
  Felice}}}\ and\ \bibinfo {author} {\bibfnamefont {F.}~\bibnamefont
  {{Sorge}}},\ }\href@noop {} {\bibfield  {journal} {\bibinfo  {journal}
  {Classical and Quantum Gravity}\ }\textbf {\bibinfo {volume} {20}},\ \bibinfo
  {pages} {469} (\bibinfo {year} {2003})}\BibitemShut {NoStop}%
\bibitem [{\citenamefont {{de Felice}}\ \emph {et~al.}(2004)\citenamefont {{de
  Felice}}, \citenamefont {{Sorge}},\ and\ \citenamefont
  {{Zilio}}}]{defelice2004}%
  \BibitemOpen
  \bibfield  {author} {\bibinfo {author} {\bibfnamefont {F.}~\bibnamefont {{de
  Felice}}}, \bibinfo {author} {\bibfnamefont {F.}~\bibnamefont {{Sorge}}}, \
  and\ \bibinfo {author} {\bibfnamefont {S.}~\bibnamefont {{Zilio}}},\ }\href
  {\doibase 10.1088/0264-9381/21/4/016} {\bibfield  {journal} {\bibinfo
  {journal} {Classical and Quantum Gravity}\ }\textbf {\bibinfo {volume}
  {21}},\ \bibinfo {pages} {961} (\bibinfo {year} {2004})}\BibitemShut
  {NoStop}%
\bibitem [{\citenamefont {{Rayimbaev}}(2016)}]{Rayimbaev16}%
  \BibitemOpen
  \bibfield  {author} {\bibinfo {author} {\bibfnamefont {J.~R.}\ \bibnamefont
  {{Rayimbaev}}},\ }\href {\doibase 10.1007/s10509-016-2879-9} {\bibfield
  {journal} {\bibinfo  {journal} {Astrophys Space Sc}\ }\textbf {\bibinfo
  {volume} {361}},\ \bibinfo {eid} {288} (\bibinfo {year} {2016})}\BibitemShut
  {NoStop}%
\bibitem [{\citenamefont {{Oteev}}\ \emph {et~al.}(2016)\citenamefont
  {{Oteev}}, \citenamefont {{Abdujabbarov}}, \citenamefont
  {{Stuchl{\'{\i}}k}},\ and\ \citenamefont {{Ahmedov}}}]{Oteev16}%
  \BibitemOpen
  \bibfield  {author} {\bibinfo {author} {\bibfnamefont {T.}~\bibnamefont
  {{Oteev}}}, \bibinfo {author} {\bibfnamefont {A.}~\bibnamefont
  {{Abdujabbarov}}}, \bibinfo {author} {\bibfnamefont {Z.}~\bibnamefont
  {{Stuchl{\'{\i}}k}}}, \ and\ \bibinfo {author} {\bibfnamefont
  {B.}~\bibnamefont {{Ahmedov}}},\ }\href {\doibase 10.1007/s10509-016-2850-9}
  {\bibfield  {journal} {\bibinfo  {journal} {Astrophys. Space Sci.}\ }\textbf
  {\bibinfo {volume} {361}},\ \bibinfo {eid} {269} (\bibinfo {year}
  {2016})}\BibitemShut {NoStop}%
\bibitem [{\citenamefont {{Toshmatov}}\ \emph {et~al.}(2015)\citenamefont
  {{Toshmatov}}, \citenamefont {{Abdujabbarov}}, \citenamefont {{Ahmedov}},\
  and\ \citenamefont {{Stuchl{\'{\i}}k}}}]{Toshmatov15d}%
  \BibitemOpen
  \bibfield  {author} {\bibinfo {author} {\bibfnamefont {B.}~\bibnamefont
  {{Toshmatov}}}, \bibinfo {author} {\bibfnamefont {A.}~\bibnamefont
  {{Abdujabbarov}}}, \bibinfo {author} {\bibfnamefont {B.}~\bibnamefont
  {{Ahmedov}}}, \ and\ \bibinfo {author} {\bibfnamefont {Z.}~\bibnamefont
  {{Stuchl{\'{\i}}k}}},\ }\href {\doibase 10.1007/s10509-015-2533-y} {\bibfield
   {journal} {\bibinfo  {journal} {Astrophys Space Sci}\ }\textbf {\bibinfo
  {volume} {360}},\ \bibinfo {eid} {19} (\bibinfo {year} {2015})}\BibitemShut
  {NoStop}%
\bibitem [{\citenamefont {{Abdujabbarov}}\ \emph {et~al.}(2014)\citenamefont
  {{Abdujabbarov}}, \citenamefont {{Ahmedov}}, \citenamefont {{Rahimov}},\ and\
  \citenamefont {{Salikhbaev}}}]{Abdujabbarov14}%
  \BibitemOpen
  \bibfield  {author} {\bibinfo {author} {\bibfnamefont {A.}~\bibnamefont
  {{Abdujabbarov}}}, \bibinfo {author} {\bibfnamefont {B.}~\bibnamefont
  {{Ahmedov}}}, \bibinfo {author} {\bibfnamefont {O.}~\bibnamefont
  {{Rahimov}}}, \ and\ \bibinfo {author} {\bibfnamefont {U.}~\bibnamefont
  {{Salikhbaev}}},\ }\href {\doibase 10.1088/0031-8949/89/8/084008} {\bibfield
  {journal} {\bibinfo  {journal} {Physica Scripta}\ }\textbf {\bibinfo {volume}
  {89}},\ \bibinfo {eid} {084008} (\bibinfo {year} {2014})}\BibitemShut
  {NoStop}%
\bibitem [{\citenamefont {{Rahimov}}\ \emph {et~al.}(2011)\citenamefont
  {{Rahimov}}, \citenamefont {{Abdujabbarov}},\ and\ \citenamefont
  {{Ahmedov}}}]{Rahimov11a}%
  \BibitemOpen
  \bibfield  {author} {\bibinfo {author} {\bibfnamefont {O.~G.}\ \bibnamefont
  {{Rahimov}}}, \bibinfo {author} {\bibfnamefont {A.~A.}\ \bibnamefont
  {{Abdujabbarov}}}, \ and\ \bibinfo {author} {\bibfnamefont {B.~J.}\
  \bibnamefont {{Ahmedov}}},\ }\href {\doibase 10.1007/s10509-011-0755-1}
  {\bibfield  {journal} {\bibinfo  {journal} {Astrophysics and Space Science}\
  }\textbf {\bibinfo {volume} {335}},\ \bibinfo {pages} {499} (\bibinfo {year}
  {2011})},\ \Eprint {http://arxiv.org/abs/1105.4543} {arXiv:1105.4543
  [astro-ph.SR]} \BibitemShut {NoStop}%
\bibitem [{\citenamefont {{Rahimov}}(2011)}]{Rahimov11}%
  \BibitemOpen
  \bibfield  {author} {\bibinfo {author} {\bibfnamefont {O.~G.}\ \bibnamefont
  {{Rahimov}}},\ }\href {\doibase 10.1142/S0217732311034931} {\bibfield
  {journal} {\bibinfo  {journal} {Modern Physics Letters A}\ }\textbf {\bibinfo
  {volume} {26}},\ \bibinfo {pages} {399} (\bibinfo {year} {2011})},\ \Eprint
  {http://arxiv.org/abs/1012.1481} {arXiv:1012.1481 [gr-qc]} \BibitemShut
  {NoStop}%
\bibitem [{\citenamefont {Haydarov}\ \emph {et~al.}(2020)\citenamefont
  {Haydarov}, \citenamefont {Abdujabbarov}, \citenamefont {Rayimbaev},\ and\
  \citenamefont {Ahmedov}}]{Haydarov20}%
  \BibitemOpen
  \bibfield  {author} {\bibinfo {author} {\bibfnamefont {K.}~\bibnamefont
  {Haydarov}}, \bibinfo {author} {\bibfnamefont {A.}~\bibnamefont
  {Abdujabbarov}}, \bibinfo {author} {\bibfnamefont {J.}~\bibnamefont
  {Rayimbaev}}, \ and\ \bibinfo {author} {\bibfnamefont {B.}~\bibnamefont
  {Ahmedov}},\ }\href {\doibase 10.3390/universe6030044} {\bibfield  {journal}
  {\bibinfo  {journal} {Universe}\ }\textbf {\bibinfo {volume} {6}} (\bibinfo
  {year} {2020}),\ 10.3390/universe6030044}\BibitemShut {NoStop}%
\bibitem [{\citenamefont {{Haydarov}}\ \emph {et~al.}(2020)\citenamefont
  {{Haydarov}}, \citenamefont {{Rayimbaev}}, \citenamefont {{Abdujabbarov}},
  \citenamefont {{Palvanov}},\ and\ \citenamefont {{Begmatova}}}]{Haydarov20b}%
  \BibitemOpen
  \bibfield  {author} {\bibinfo {author} {\bibfnamefont {K.}~\bibnamefont
  {{Haydarov}}}, \bibinfo {author} {\bibfnamefont {J.}~\bibnamefont
  {{Rayimbaev}}}, \bibinfo {author} {\bibfnamefont {A.}~\bibnamefont
  {{Abdujabbarov}}}, \bibinfo {author} {\bibfnamefont {S.}~\bibnamefont
  {{Palvanov}}}, \ and\ \bibinfo {author} {\bibfnamefont {D.}~\bibnamefont
  {{Begmatova}}},\ }\href@noop {} {\bibfield  {journal} {\bibinfo  {journal}
  {arXiv e-prints}\ ,\ \bibinfo {eid} {arXiv:2004.14868}} (\bibinfo {year}
  {2020})},\ \Eprint {http://arxiv.org/abs/arXiv:2004.14868}
  {arXiv:arXiv:2004.14868 [gr-qc]} \BibitemShut {NoStop}%
\bibitem [{\citenamefont {{Rayimbaev}}\ \emph
  {et~al.}(2020{\natexlab{a}})\citenamefont {{Rayimbaev}}, \citenamefont
  {{Abdujabbarov}}, \citenamefont {{Turimov}},\ and\ \citenamefont
  {{Atamurotov}}}]{Rayimbaev20204DEGB}%
  \BibitemOpen
  \bibfield  {author} {\bibinfo {author} {\bibfnamefont {J.}~\bibnamefont
  {{Rayimbaev}}}, \bibinfo {author} {\bibfnamefont {A.}~\bibnamefont
  {{Abdujabbarov}}}, \bibinfo {author} {\bibfnamefont {B.}~\bibnamefont
  {{Turimov}}}, \ and\ \bibinfo {author} {\bibfnamefont {F.}~\bibnamefont
  {{Atamurotov}}},\ }\href@noop {} {\bibfield  {journal} {\bibinfo  {journal}
  {arXiv e-prints}\ ,\ \bibinfo {eid} {arXiv:2004.10031}} (\bibinfo {year}
  {2020}{\natexlab{a}})},\ \Eprint {http://arxiv.org/abs/2004.10031}
  {arXiv:2004.10031 [gr-qc]} \BibitemShut {NoStop}%
\bibitem [{\citenamefont {{Kov{\'a}{\v r}}}\ \emph {et~al.}(2010)\citenamefont
  {{Kov{\'a}{\v r}}}, \citenamefont {{Kop{\'a}{\v c}ek}}, \citenamefont
  {{Karas}},\ and\ \citenamefont {{Stuchl{\'{\i}}k}}}]{Kovar10}%
  \BibitemOpen
  \bibfield  {author} {\bibinfo {author} {\bibfnamefont {J.}~\bibnamefont
  {{Kov{\'a}{\v r}}}}, \bibinfo {author} {\bibfnamefont {O.}~\bibnamefont
  {{Kop{\'a}{\v c}ek}}}, \bibinfo {author} {\bibfnamefont {V.}~\bibnamefont
  {{Karas}}}, \ and\ \bibinfo {author} {\bibfnamefont {Z.}~\bibnamefont
  {{Stuchl{\'{\i}}k}}},\ }\href {\doibase 10.1088/0264-9381/27/13/135006}
  {\bibfield  {journal} {\bibinfo  {journal} {Classical and Quantum Gravity}\
  }\textbf {\bibinfo {volume} {27}},\ \bibinfo {eid} {135006} (\bibinfo {year}
  {2010})},\ \Eprint {http://arxiv.org/abs/1005.3270} {arXiv:1005.3270
  [astro-ph.HE]} \BibitemShut {NoStop}%
\bibitem [{\citenamefont {{Kov{\'a}{\v r}}}\ \emph {et~al.}(2014)\citenamefont
  {{Kov{\'a}{\v r}}}, \citenamefont {{Slan{\'y}}}, \citenamefont
  {{Cremaschini}}, \citenamefont {{Stuchl{\'{\i}}k}}, \citenamefont {{Karas}},\
  and\ \citenamefont {{Trova}}}]{Kovar14}%
  \BibitemOpen
  \bibfield  {author} {\bibinfo {author} {\bibfnamefont {J.}~\bibnamefont
  {{Kov{\'a}{\v r}}}}, \bibinfo {author} {\bibfnamefont {P.}~\bibnamefont
  {{Slan{\'y}}}}, \bibinfo {author} {\bibfnamefont {C.}~\bibnamefont
  {{Cremaschini}}}, \bibinfo {author} {\bibfnamefont {Z.}~\bibnamefont
  {{Stuchl{\'{\i}}k}}}, \bibinfo {author} {\bibfnamefont {V.}~\bibnamefont
  {{Karas}}}, \ and\ \bibinfo {author} {\bibfnamefont {A.}~\bibnamefont
  {{Trova}}},\ }\href {\doibase 10.1103/PhysRevD.90.044029} {\bibfield
  {journal} {\bibinfo  {journal} {Phys. Rev. D}\ }\textbf {\bibinfo {volume}
  {90}},\ \bibinfo {eid} {044029} (\bibinfo {year} {2014})},\ \Eprint
  {http://arxiv.org/abs/1409.0418} {arXiv:1409.0418 [gr-qc]} \BibitemShut
  {NoStop}%
\bibitem [{\citenamefont {{Aliev}}\ and\ \citenamefont
  {{Gal'tsov}}(1989)}]{Aliev89}%
  \BibitemOpen
  \bibfield  {author} {\bibinfo {author} {\bibfnamefont {A.~N.}\ \bibnamefont
  {{Aliev}}}\ and\ \bibinfo {author} {\bibfnamefont {D.~V.}\ \bibnamefont
  {{Gal'tsov}}},\ }\href {\doibase 10.1070/PU1989v032n01ABEH002677} {\bibfield
  {journal} {\bibinfo  {journal} {Soviet Physics Uspekhi}\ }\textbf {\bibinfo
  {volume} {32}},\ \bibinfo {pages} {75} (\bibinfo {year} {1989})}\BibitemShut
  {NoStop}%
\bibitem [{\citenamefont {{Aliev}}\ and\ \citenamefont
  {{{\"O}zdemir}}(2002)}]{Aliev02}%
  \BibitemOpen
  \bibfield  {author} {\bibinfo {author} {\bibfnamefont {A.~N.}\ \bibnamefont
  {{Aliev}}}\ and\ \bibinfo {author} {\bibfnamefont {N.}~\bibnamefont
  {{{\"O}zdemir}}},\ }\href {\doibase 10.1046/j.1365-8711.2002.05727.x}
  {\bibfield  {journal} {\bibinfo  {journal} {Mon. Not. R. Astron. Soc.}\
  }\textbf {\bibinfo {volume} {336}},\ \bibinfo {pages} {241} (\bibinfo {year}
  {2002})},\ \Eprint {http://arxiv.org/abs/gr-qc/0208025} {gr-qc/0208025}
  \BibitemShut {NoStop}%
\bibitem [{\citenamefont {{Aliev}}\ \emph {et~al.}(1986)\citenamefont
  {{Aliev}}, \citenamefont {{Galtsov}},\ and\ \citenamefont
  {{Petukhov}}}]{Aliev86}%
  \BibitemOpen
  \bibfield  {author} {\bibinfo {author} {\bibfnamefont {A.~N.}\ \bibnamefont
  {{Aliev}}}, \bibinfo {author} {\bibfnamefont {D.~V.}\ \bibnamefont
  {{Galtsov}}}, \ and\ \bibinfo {author} {\bibfnamefont {V.~I.}\ \bibnamefont
  {{Petukhov}}},\ }\href {\doibase 10.1007/BF00649756} {\bibfield  {journal}
  {\bibinfo  {journal} {Astrophys. Space Sci.}\ }\textbf {\bibinfo {volume}
  {124}},\ \bibinfo {pages} {137} (\bibinfo {year} {1986})}\BibitemShut
  {NoStop}%
\bibitem [{\citenamefont {{Frolov}}\ and\ \citenamefont
  {{Krtou{\v{s}}}}(2011)}]{Frolov11}%
  \BibitemOpen
  \bibfield  {author} {\bibinfo {author} {\bibfnamefont {V.~P.}\ \bibnamefont
  {{Frolov}}}\ and\ \bibinfo {author} {\bibfnamefont {P.}~\bibnamefont
  {{Krtou{\v{s}}}}},\ }\href {\doibase 10.1103/PhysRevD.83.024016} {\bibfield
  {journal} {\bibinfo  {journal} {Phys. Rev. D}\ }\textbf {\bibinfo {volume}
  {83}},\ \bibinfo {eid} {024016} (\bibinfo {year} {2011})},\ \Eprint
  {http://arxiv.org/abs/1010.2266} {arXiv:1010.2266 [hep-th]} \BibitemShut
  {NoStop}%
\bibitem [{\citenamefont {{Frolov}}(2012)}]{Frolov12}%
  \BibitemOpen
  \bibfield  {author} {\bibinfo {author} {\bibfnamefont {V.~P.}\ \bibnamefont
  {{Frolov}}},\ }\href {\doibase 10.1103/PhysRevD.85.024020} {\bibfield
  {journal} {\bibinfo  {journal} {Phys. Rev. D.}\ }\textbf {\bibinfo {volume}
  {85}},\ \bibinfo {eid} {024020} (\bibinfo {year} {2012})},\ \Eprint
  {http://arxiv.org/abs/1110.6274} {arXiv:1110.6274 [gr-qc]} \BibitemShut
  {NoStop}%
\bibitem [{\citenamefont {{Stuchl{\'{\i}}k}}\ \emph {et~al.}(2014)\citenamefont
  {{Stuchl{\'{\i}}k}}, \citenamefont {{Schee}},\ and\ \citenamefont
  {{Abdujabbarov}}}]{Stuchlik14a}%
  \BibitemOpen
  \bibfield  {author} {\bibinfo {author} {\bibfnamefont {Z.}~\bibnamefont
  {{Stuchl{\'{\i}}k}}}, \bibinfo {author} {\bibfnamefont {J.}~\bibnamefont
  {{Schee}}}, \ and\ \bibinfo {author} {\bibfnamefont {A.}~\bibnamefont
  {{Abdujabbarov}}},\ }\href {\doibase 10.1103/PhysRevD.89.104048} {\bibfield
  {journal} {\bibinfo  {journal} {Phys. Rev. D}\ }\textbf {\bibinfo {volume}
  {89}},\ \bibinfo {eid} {104048} (\bibinfo {year} {2014})}\BibitemShut
  {NoStop}%
\bibitem [{\citenamefont {{Shaymatov}}\ \emph {et~al.}(2014)\citenamefont
  {{Shaymatov}}, \citenamefont {{Atamurotov}},\ and\ \citenamefont
  {{Ahmedov}}}]{Shaymatov14}%
  \BibitemOpen
  \bibfield  {author} {\bibinfo {author} {\bibfnamefont {S.}~\bibnamefont
  {{Shaymatov}}}, \bibinfo {author} {\bibfnamefont {F.}~\bibnamefont
  {{Atamurotov}}}, \ and\ \bibinfo {author} {\bibfnamefont {B.}~\bibnamefont
  {{Ahmedov}}},\ }\href {\doibase 10.1007/s10509-013-1752-3} {\bibfield
  {journal} {\bibinfo  {journal} {Astrophys Space Sci}\ }\textbf {\bibinfo
  {volume} {350}},\ \bibinfo {pages} {413} (\bibinfo {year}
  {2014})}\BibitemShut {NoStop}%
\bibitem [{\citenamefont {{Abdujabbarov}}\ and\ \citenamefont
  {{Ahmedov}}(2010)}]{Abdujabbarov10}%
  \BibitemOpen
  \bibfield  {author} {\bibinfo {author} {\bibfnamefont {A.}~\bibnamefont
  {{Abdujabbarov}}}\ and\ \bibinfo {author} {\bibfnamefont {B.}~\bibnamefont
  {{Ahmedov}}},\ }\href {\doibase 10.1103/PhysRevD.81.044022} {\bibfield
  {journal} {\bibinfo  {journal} {Phys. Rev. D}\ }\textbf {\bibinfo {volume}
  {81}},\ \bibinfo {eid} {044022} (\bibinfo {year} {2010})},\ \Eprint
  {http://arxiv.org/abs/0905.2730} {arXiv:0905.2730 [gr-qc]} \BibitemShut
  {NoStop}%
\bibitem [{\citenamefont {{Abdujabbarov}}\ \emph
  {et~al.}(2011{\natexlab{a}})\citenamefont {{Abdujabbarov}}, \citenamefont
  {{Ahmedov}},\ and\ \citenamefont {{Hakimov}}}]{Abdujabbarov11a}%
  \BibitemOpen
  \bibfield  {author} {\bibinfo {author} {\bibfnamefont {A.}~\bibnamefont
  {{Abdujabbarov}}}, \bibinfo {author} {\bibfnamefont {B.}~\bibnamefont
  {{Ahmedov}}}, \ and\ \bibinfo {author} {\bibfnamefont {A.}~\bibnamefont
  {{Hakimov}}},\ }\href {\doibase 10.1103/PhysRevD.83.044053} {\bibfield
  {journal} {\bibinfo  {journal} {Phys.Rev. D}\ }\textbf {\bibinfo {volume}
  {83}},\ \bibinfo {eid} {044053} (\bibinfo {year} {2011}{\natexlab{a}})},\
  \Eprint {http://arxiv.org/abs/1101.4741} {arXiv:1101.4741 [gr-qc]}
  \BibitemShut {NoStop}%
\bibitem [{\citenamefont {{Abdujabbarov}}\ \emph
  {et~al.}(2011{\natexlab{b}})\citenamefont {{Abdujabbarov}}, \citenamefont
  {{Ahmedov}}, \citenamefont {{Shaymatov}},\ and\ \citenamefont
  {{Rakhmatov}}}]{Abdujabbarov11}%
  \BibitemOpen
  \bibfield  {author} {\bibinfo {author} {\bibfnamefont {A.~A.}\ \bibnamefont
  {{Abdujabbarov}}}, \bibinfo {author} {\bibfnamefont {B.~J.}\ \bibnamefont
  {{Ahmedov}}}, \bibinfo {author} {\bibfnamefont {S.~R.}\ \bibnamefont
  {{Shaymatov}}}, \ and\ \bibinfo {author} {\bibfnamefont {A.~S.}\ \bibnamefont
  {{Rakhmatov}}},\ }\href {\doibase 10.1007/s10509-011-0740-8} {\bibfield
  {journal} {\bibinfo  {journal} {Astrophys Space Sci}\ }\textbf {\bibinfo
  {volume} {334}},\ \bibinfo {pages} {237} (\bibinfo {year}
  {2011}{\natexlab{b}})},\ \Eprint {http://arxiv.org/abs/1105.1910}
  {arXiv:1105.1910 [astro-ph.SR]} \BibitemShut {NoStop}%
\bibitem [{\citenamefont {{Abdujabbarov}}\ \emph {et~al.}(2008)\citenamefont
  {{Abdujabbarov}}, \citenamefont {{Ahmedov}},\ and\ \citenamefont
  {{Kagramanova}}}]{Abdujabbarov08}%
  \BibitemOpen
  \bibfield  {author} {\bibinfo {author} {\bibfnamefont {A.~A.}\ \bibnamefont
  {{Abdujabbarov}}}, \bibinfo {author} {\bibfnamefont {B.~J.}\ \bibnamefont
  {{Ahmedov}}}, \ and\ \bibinfo {author} {\bibfnamefont {V.~G.}\ \bibnamefont
  {{Kagramanova}}},\ }\href {\doibase 10.1007/s10714-008-0635-3} {\bibfield
  {journal} {\bibinfo  {journal} {General Relativity and Gravitation}\ }\textbf
  {\bibinfo {volume} {40}},\ \bibinfo {pages} {2515} (\bibinfo {year}
  {2008})},\ \Eprint {http://arxiv.org/abs/0802.4349} {arXiv:0802.4349 [gr-qc]}
  \BibitemShut {NoStop}%
\bibitem [{\citenamefont {{Karas}}\ \emph {et~al.}(2012)\citenamefont
  {{Karas}}, \citenamefont {{Kovar}}, \citenamefont {{Kopacek}}, \citenamefont
  {{Kojima}}, \citenamefont {{Slany}},\ and\ \citenamefont
  {{Stuchlik}}}]{Karas12a}%
  \BibitemOpen
  \bibfield  {author} {\bibinfo {author} {\bibfnamefont {V.}~\bibnamefont
  {{Karas}}}, \bibinfo {author} {\bibfnamefont {J.}~\bibnamefont {{Kovar}}},
  \bibinfo {author} {\bibfnamefont {O.}~\bibnamefont {{Kopacek}}}, \bibinfo
  {author} {\bibfnamefont {Y.}~\bibnamefont {{Kojima}}}, \bibinfo {author}
  {\bibfnamefont {P.}~\bibnamefont {{Slany}}}, \ and\ \bibinfo {author}
  {\bibfnamefont {Z.}~\bibnamefont {{Stuchlik}}},\ }in\ \href@noop {} {\emph
  {\bibinfo {booktitle} {American Astronomical Society Meeting Abstracts
  \#220}}},\ \bibinfo {series} {American Astronomical Society Meeting
  Abstracts}, Vol.\ \bibinfo {volume} {220}\ (\bibinfo {year} {2012})\ p.\
  \bibinfo {pages} {430.07}\BibitemShut {NoStop}%
\bibitem [{\citenamefont {{Shaymatov}}\ \emph {et~al.}(2015)\citenamefont
  {{Shaymatov}}, \citenamefont {{Patil}}, \citenamefont {{Ahmedov}},\ and\
  \citenamefont {{Joshi}}}]{Shaymatov15}%
  \BibitemOpen
  \bibfield  {author} {\bibinfo {author} {\bibfnamefont {S.}~\bibnamefont
  {{Shaymatov}}}, \bibinfo {author} {\bibfnamefont {M.}~\bibnamefont
  {{Patil}}}, \bibinfo {author} {\bibfnamefont {B.}~\bibnamefont {{Ahmedov}}},
  \ and\ \bibinfo {author} {\bibfnamefont {P.~S.}\ \bibnamefont {{Joshi}}},\
  }\href {\doibase 10.1103/PhysRevD.91.064025} {\bibfield  {journal} {\bibinfo
  {journal} {Phys. Rev. D}\ }\textbf {\bibinfo {volume} {91}},\ \bibinfo {eid}
  {064025} (\bibinfo {year} {2015})},\ \Eprint {http://arxiv.org/abs/1409.3018}
  {arXiv:1409.3018 [gr-qc]} \BibitemShut {NoStop}%
\bibitem [{\citenamefont {{Stuchl{\'{\i}}k}}\ and\ \citenamefont {{Kolo{\v
  s}}}(2016)}]{Stuchlik16}%
  \BibitemOpen
  \bibfield  {author} {\bibinfo {author} {\bibfnamefont {Z.}~\bibnamefont
  {{Stuchl{\'{\i}}k}}}\ and\ \bibinfo {author} {\bibfnamefont {M.}~\bibnamefont
  {{Kolo{\v s}}}},\ }\href {\doibase 10.1140/epjc/s10052-015-3862-2} {\bibfield
   {journal} {\bibinfo  {journal} {European Physical Journal C}\ }\textbf
  {\bibinfo {volume} {76}},\ \bibinfo {eid} {32} (\bibinfo {year} {2016})},\
  \Eprint {http://arxiv.org/abs/1511.02936} {arXiv:1511.02936 [gr-qc]}
  \BibitemShut {NoStop}%
\bibitem [{\citenamefont {{Rayimbaev}}\ \emph
  {et~al.}(2020{\natexlab{b}})\citenamefont {{Rayimbaev}}, \citenamefont
  {{Turimov}}, \citenamefont {{Marcos}}, \citenamefont {{Palvanov}},\ and\
  \citenamefont {{Rakhmatov}}}]{Rayimbaev20}%
  \BibitemOpen
  \bibfield  {author} {\bibinfo {author} {\bibfnamefont {J.}~\bibnamefont
  {{Rayimbaev}}}, \bibinfo {author} {\bibfnamefont {B.}~\bibnamefont
  {{Turimov}}}, \bibinfo {author} {\bibfnamefont {F.}~\bibnamefont {{Marcos}}},
  \bibinfo {author} {\bibfnamefont {S.}~\bibnamefont {{Palvanov}}}, \ and\
  \bibinfo {author} {\bibfnamefont {A.}~\bibnamefont {{Rakhmatov}}},\ }\href
  {\doibase 10.1142/S021773232050056X} {\bibfield  {journal} {\bibinfo
  {journal} {Modern Physics Letters A}\ }\textbf {\bibinfo {volume} {35}},\
  \bibinfo {eid} {2050056} (\bibinfo {year} {2020}{\natexlab{b}})}\BibitemShut
  {NoStop}%
\bibitem [{\citenamefont {{Turimov}}(2018)}]{Turimov18IJMPD}%
  \BibitemOpen
  \bibfield  {author} {\bibinfo {author} {\bibfnamefont {B.}~\bibnamefont
  {{Turimov}}},\ }\href {\doibase 10.1142/S021827181850092X} {\bibfield
  {journal} {\bibinfo  {journal} {International Journal of Modern Physics D}\
  }\textbf {\bibinfo {volume} {27}},\ \bibinfo {eid} {1850092} (\bibinfo {year}
  {2018})}\BibitemShut {NoStop}%
\bibitem [{\citenamefont {{Turimov}}\ \emph {et~al.}(2018)\citenamefont
  {{Turimov}}, \citenamefont {{Ahmedov}}, \citenamefont {{Abdujabbarov}},\ and\
  \citenamefont {{Bambi}}}]{Turimov18b}%
  \BibitemOpen
  \bibfield  {author} {\bibinfo {author} {\bibfnamefont {B.}~\bibnamefont
  {{Turimov}}}, \bibinfo {author} {\bibfnamefont {B.}~\bibnamefont
  {{Ahmedov}}}, \bibinfo {author} {\bibfnamefont {A.}~\bibnamefont
  {{Abdujabbarov}}}, \ and\ \bibinfo {author} {\bibfnamefont {C.}~\bibnamefont
  {{Bambi}}},\ }\href {\doibase 10.1103/PhysRevD.97.124005} {\bibfield
  {journal} {\bibinfo  {journal} {Phys. Rev. D}\ }\textbf {\bibinfo {volume}
  {97}},\ \bibinfo {eid} {124005} (\bibinfo {year} {2018})},\ \Eprint
  {http://arxiv.org/abs/1805.00005} {arXiv:1805.00005 [gr-qc]} \BibitemShut
  {NoStop}%
\bibitem [{\citenamefont {{Shaymatov}}\ \emph {et~al.}(2018)\citenamefont
  {{Shaymatov}}, \citenamefont {{Ahmedov}}, \citenamefont {{Stuchl{\'{\i}}k}},\
  and\ \citenamefont {{Abdujabbarov}}}]{Shaymatov18}%
  \BibitemOpen
  \bibfield  {author} {\bibinfo {author} {\bibfnamefont {S.}~\bibnamefont
  {{Shaymatov}}}, \bibinfo {author} {\bibfnamefont {B.}~\bibnamefont
  {{Ahmedov}}}, \bibinfo {author} {\bibfnamefont {Z.}~\bibnamefont
  {{Stuchl{\'{\i}}k}}}, \ and\ \bibinfo {author} {\bibfnamefont
  {A.}~\bibnamefont {{Abdujabbarov}}},\ }\href {\doibase
  10.1142/S0218271818500888} {\bibfield  {journal} {\bibinfo  {journal}
  {International Journal of Modern Physics D}\ }\textbf {\bibinfo {volume}
  {27}},\ \bibinfo {eid} {1850088} (\bibinfo {year} {2018})}\BibitemShut
  {NoStop}%
\bibitem [{\citenamefont {{Turimov}}\ \emph {et~al.}(2017)\citenamefont
  {{Turimov}}, \citenamefont {{Ahmedov}},\ and\ \citenamefont
  {{Hakimov}}}]{Turimov17}%
  \BibitemOpen
  \bibfield  {author} {\bibinfo {author} {\bibfnamefont {B.~V.}\ \bibnamefont
  {{Turimov}}}, \bibinfo {author} {\bibfnamefont {B.~J.}\ \bibnamefont
  {{Ahmedov}}}, \ and\ \bibinfo {author} {\bibfnamefont {A.~A.}\ \bibnamefont
  {{Hakimov}}},\ }\href {\doibase 10.1103/PhysRevD.96.104001} {\bibfield
  {journal} {\bibinfo  {journal} {Phys. Rev. D}\ }\textbf {\bibinfo {volume}
  {96}},\ \bibinfo {eid} {104001} (\bibinfo {year} {2017})}\BibitemShut
  {NoStop}%
\bibitem [{\citenamefont {{Shaymatov}}\ \emph
  {et~al.}(2020{\natexlab{a}})\citenamefont {{Shaymatov}}, \citenamefont
  {{Vrba}}, \citenamefont {{Malafarina}}, \citenamefont {{Ahmedov}},\ and\
  \citenamefont {{Stuchl{\'\i}k}}}]{Shaymatov20egb}%
  \BibitemOpen
  \bibfield  {author} {\bibinfo {author} {\bibfnamefont {S.}~\bibnamefont
  {{Shaymatov}}}, \bibinfo {author} {\bibfnamefont {J.}~\bibnamefont {{Vrba}}},
  \bibinfo {author} {\bibfnamefont {D.}~\bibnamefont {{Malafarina}}}, \bibinfo
  {author} {\bibfnamefont {B.}~\bibnamefont {{Ahmedov}}}, \ and\ \bibinfo
  {author} {\bibfnamefont {Z.}~\bibnamefont {{Stuchl{\'\i}k}}},\ }\href@noop {}
  {\bibfield  {journal} {\bibinfo  {journal} {arXiv e-prints}\ ,\ \bibinfo
  {eid} {arXiv:2005.12410}} (\bibinfo {year} {2020}{\natexlab{a}})},\ \Eprint
  {http://arxiv.org/abs/2005.12410} {arXiv:2005.12410 [gr-qc]} \BibitemShut
  {NoStop}%
\bibitem [{\citenamefont {{Rayimbaev}}\ \emph {et~al.}(2015)\citenamefont
  {{Rayimbaev}}, \citenamefont {{Ahmedov}}, \citenamefont {{Juraeva}},\ and\
  \citenamefont {{Rakhmatov}}}]{Rayimbaev15}%
  \BibitemOpen
  \bibfield  {author} {\bibinfo {author} {\bibfnamefont {J.~R.}\ \bibnamefont
  {{Rayimbaev}}}, \bibinfo {author} {\bibfnamefont {B.~J.}\ \bibnamefont
  {{Ahmedov}}}, \bibinfo {author} {\bibfnamefont {N.~B.}\ \bibnamefont
  {{Juraeva}}}, \ and\ \bibinfo {author} {\bibfnamefont {A.~S.}\ \bibnamefont
  {{Rakhmatov}}},\ }\href {\doibase 10.1007/s10509-014-2208-0} {\bibfield
  {journal} {\bibinfo  {journal} {Astrophys Space Sc}\ }\textbf {\bibinfo
  {volume} {356}},\ \bibinfo {pages} {301} (\bibinfo {year}
  {2015})}\BibitemShut {NoStop}%
\bibitem [{\citenamefont {{Shaymatov}}(2019)}]{shaymatov19b}%
  \BibitemOpen
  \bibfield  {author} {\bibinfo {author} {\bibfnamefont {S.}~\bibnamefont
  {{Shaymatov}}},\ }\href {\doibase 10.1142/S2010194519600206} {\bibfield
  {journal} {\bibinfo  {journal} {Int. J. Mod. Phys. Conf. Ser.}\ }\textbf
  {\bibinfo {volume} {49}},\ \bibinfo {pages} {1960020} (\bibinfo {year}
  {2019})}\BibitemShut {NoStop}%
\bibitem [{\citenamefont {{Rayimbaev}}\ \emph {et~al.}(2019)\citenamefont
  {{Rayimbaev}}, \citenamefont {{Turimov}},\ and\ \citenamefont
  {{Ahmedov}}}]{Rayimbaev19}%
  \BibitemOpen
  \bibfield  {author} {\bibinfo {author} {\bibfnamefont {J.}~\bibnamefont
  {{Rayimbaev}}}, \bibinfo {author} {\bibfnamefont {B.}~\bibnamefont
  {{Turimov}}}, \ and\ \bibinfo {author} {\bibfnamefont {B.}~\bibnamefont
  {{Ahmedov}}},\ }\href {\doibase 10.1142/S0218271819501281} {\bibfield
  {journal} {\bibinfo  {journal} {International Journal of Modern Physics D}\
  }\textbf {\bibinfo {volume} {28}},\ \bibinfo {eid} {1950128-209} (\bibinfo
  {year} {2019})}\BibitemShut {NoStop}%
\bibitem [{\citenamefont {{Shaymatov}}\ \emph
  {et~al.}(2020{\natexlab{b}})\citenamefont {{Shaymatov}}, \citenamefont
  {{Malafarina}},\ and\ \citenamefont {{Ahmedov}}}]{Shaymatov20b}%
  \BibitemOpen
  \bibfield  {author} {\bibinfo {author} {\bibfnamefont {S.}~\bibnamefont
  {{Shaymatov}}}, \bibinfo {author} {\bibfnamefont {D.}~\bibnamefont
  {{Malafarina}}}, \ and\ \bibinfo {author} {\bibfnamefont {B.}~\bibnamefont
  {{Ahmedov}}},\ }\href@noop {} {\bibfield  {journal} {\bibinfo  {journal}
  {arXiv e-prints}\ ,\ \bibinfo {eid} {arXiv:2004.06811}} (\bibinfo {year}
  {2020}{\natexlab{b}})},\ \Eprint {http://arxiv.org/abs/2004.06811}
  {arXiv:2004.06811 [gr-qc]} \BibitemShut {NoStop}%
\bibitem [{\citenamefont {{Narzilloev}}\ \emph {et~al.}(2020)\citenamefont
  {{Narzilloev}}, \citenamefont {{Rayimbaev}}, \citenamefont {{Abdujabbarov}},\
  and\ \citenamefont {{Bambi}}}]{Narzilloev2020C}%
  \BibitemOpen
  \bibfield  {author} {\bibinfo {author} {\bibfnamefont {B.}~\bibnamefont
  {{Narzilloev}}}, \bibinfo {author} {\bibfnamefont {J.}~\bibnamefont
  {{Rayimbaev}}}, \bibinfo {author} {\bibfnamefont {A.}~\bibnamefont
  {{Abdujabbarov}}}, \ and\ \bibinfo {author} {\bibfnamefont {C.}~\bibnamefont
  {{Bambi}}},\ }\href@noop {} {\bibfield  {journal} {\bibinfo  {journal} {arXiv
  e-prints}\ ,\ \bibinfo {eid} {arXiv:2005.04752}} (\bibinfo {year} {2020})},\
  \Eprint {http://arxiv.org/abs/2005.04752} {arXiv:2005.04752 [gr-qc]}
  \BibitemShut {NoStop}%
\bibitem [{\citenamefont {{Narzilloev}}\ \emph {et~al.}(2019)\citenamefont
  {{Narzilloev}}, \citenamefont {{Abdujabbarov}}, \citenamefont {{Bambi}},\
  and\ \citenamefont {{Ahmedov}}}]{Narzilloev19}%
  \BibitemOpen
  \bibfield  {author} {\bibinfo {author} {\bibfnamefont {B.}~\bibnamefont
  {{Narzilloev}}}, \bibinfo {author} {\bibfnamefont {A.}~\bibnamefont
  {{Abdujabbarov}}}, \bibinfo {author} {\bibfnamefont {C.}~\bibnamefont
  {{Bambi}}}, \ and\ \bibinfo {author} {\bibfnamefont {B.}~\bibnamefont
  {{Ahmedov}}},\ }\href {\doibase 10.1103/PhysRevD.99.104009} {\bibfield
  {journal} {\bibinfo  {journal} {Phys. Rev. D}\ }\textbf {\bibinfo {volume}
  {99}},\ \bibinfo {eid} {104009} (\bibinfo {year} {2019})},\ \Eprint
  {http://arxiv.org/abs/1902.03414} {arXiv:1902.03414 [gr-qc]} \BibitemShut
  {NoStop}%
\bibitem [{\citenamefont {Rayimbaev}\ \emph {et~al.}(2020)\citenamefont
  {Rayimbaev}, \citenamefont {Figueroa}, \citenamefont {Stuchl\'{\i}k},\ and\
  \citenamefont {Juraev}}]{Rayimbaev2020PRD}%
  \BibitemOpen
  \bibfield  {author} {\bibinfo {author} {\bibfnamefont {J.}~\bibnamefont
  {Rayimbaev}}, \bibinfo {author} {\bibfnamefont {M.}~\bibnamefont {Figueroa}},
  \bibinfo {author} {\bibfnamefont {Z.~c.~v.}\ \bibnamefont {Stuchl\'{\i}k}}, \
  and\ \bibinfo {author} {\bibfnamefont {B.}~\bibnamefont {Juraev}},\ }\href
  {\doibase 10.1103/PhysRevD.101.104045} {\bibfield  {journal} {\bibinfo
  {journal} {Phys. Rev. D}\ }\textbf {\bibinfo {volume} {101}},\ \bibinfo
  {pages} {104045} (\bibinfo {year} {2020})}\BibitemShut {NoStop}%
\bibitem [{\citenamefont {{Yu}}\ \emph {et~al.}(2020)\citenamefont {{Yu}},
  \citenamefont {{Qiu}},\ and\ \citenamefont {{Gao}}}]{Yu20}%
  \BibitemOpen
  \bibfield  {author} {\bibinfo {author} {\bibfnamefont {S.}~\bibnamefont
  {{Yu}}}, \bibinfo {author} {\bibfnamefont {J.}~\bibnamefont {{Qiu}}}, \ and\
  \bibinfo {author} {\bibfnamefont {C.}~\bibnamefont {{Gao}}},\ }\href@noop {}
  {\bibfield  {journal} {\bibinfo  {journal} {arXiv e-prints}\ ,\ \bibinfo
  {eid} {arXiv:2005.14476}} (\bibinfo {year} {2020})},\ \Eprint
  {http://arxiv.org/abs/2005.14476} {arXiv:2005.14476 [gr-qc]} \BibitemShut
  {NoStop}%
\bibitem [{\citenamefont {{Gao}}\ and\ \citenamefont {{Zhang}}(2004)}]{Gao04}%
  \BibitemOpen
  \bibfield  {author} {\bibinfo {author} {\bibfnamefont {C.~J.}\ \bibnamefont
  {{Gao}}}\ and\ \bibinfo {author} {\bibfnamefont {S.~N.}\ \bibnamefont
  {{Zhang}}},\ }\href {\doibase 10.1103/PhysRevD.70.124019} {\bibfield
  {journal} {\bibinfo  {journal} {Phys. Rev. D}\ }\textbf {\bibinfo {volume}
  {70}},\ \bibinfo {eid} {124019} (\bibinfo {year} {2004})},\ \Eprint
  {http://arxiv.org/abs/hep-th/0411104} {arXiv:hep-th/0411104 [astro-ph]}
  \BibitemShut {NoStop}%
\bibitem [{\citenamefont {{Stuchlik}}(1983)}]{Stuchlik83}%
  \BibitemOpen
  \bibfield  {author} {\bibinfo {author} {\bibfnamefont {Z.}~\bibnamefont
  {{Stuchlik}}},\ }\href@noop {} {\bibfield  {journal} {\bibinfo  {journal}
  {Bulletin of the Astronomical Institutes of Czechoslovakia}\ }\textbf
  {\bibinfo {volume} {34}},\ \bibinfo {pages} {129} (\bibinfo {year}
  {1983})}\BibitemShut {NoStop}%
\bibitem [{\citenamefont {{Stuchl{\'{\i}}k}}\ and\ \citenamefont
  {{Hled{\'{\i}}k}}(1999)}]{Stuchlik99a}%
  \BibitemOpen
  \bibfield  {author} {\bibinfo {author} {\bibfnamefont {Z.}~\bibnamefont
  {{Stuchl{\'{\i}}k}}}\ and\ \bibinfo {author} {\bibfnamefont {S.}~\bibnamefont
  {{Hled{\'{\i}}k}}},\ }\href {\doibase 10.1103/PhysRevD.60.044006} {\bibfield
  {journal} {\bibinfo  {journal} {Phys. Rev. D}\ }\textbf {\bibinfo {volume}
  {60}},\ \bibinfo {eid} {044006} (\bibinfo {year} {1999})}\BibitemShut
  {NoStop}%
\bibitem [{\citenamefont {{Stuchl{\'{\i}}k}}\ and\ \citenamefont
  {{Hled{\'{\i}}k}}(2000)}]{Stuchlik00a}%
  \BibitemOpen
  \bibfield  {author} {\bibinfo {author} {\bibfnamefont {Z.}~\bibnamefont
  {{Stuchl{\'{\i}}k}}}\ and\ \bibinfo {author} {\bibfnamefont {S.}~\bibnamefont
  {{Hled{\'{\i}}k}}},\ }\href {\doibase 10.1088/0264-9381/17/21/312} {\bibfield
   {journal} {\bibinfo  {journal} {Classical and Quantum Gravity}\ }\textbf
  {\bibinfo {volume} {17}},\ \bibinfo {pages} {4541} (\bibinfo {year}
  {2000})},\ \Eprint {http://arxiv.org/abs/0803.2539} {arXiv:0803.2539 [gr-qc]}
  \BibitemShut {NoStop}%
\bibitem [{\citenamefont {{Stuchl{\'{\i}}k}}\ and\ \citenamefont
  {{Hledik}}(2002)}]{Stuchlik02}%
  \BibitemOpen
  \bibfield  {author} {\bibinfo {author} {\bibfnamefont {Z.}~\bibnamefont
  {{Stuchl{\'{\i}}k}}}\ and\ \bibinfo {author} {\bibfnamefont {S.}~\bibnamefont
  {{Hledik}}},\ }\href@noop {} {\bibfield  {journal} {\bibinfo  {journal} {Acta
  Physica Slovaca}\ }\textbf {\bibinfo {volume} {52}},\ \bibinfo {pages} {363}
  (\bibinfo {year} {2002})}\BibitemShut {NoStop}%
\bibitem [{\citenamefont {{Stuchl{\'{\i}}k}}\ and\ \citenamefont
  {{Slan{\'y}}}(2004)}]{Stuchlik04}%
  \BibitemOpen
  \bibfield  {author} {\bibinfo {author} {\bibfnamefont {Z.}~\bibnamefont
  {{Stuchl{\'{\i}}k}}}\ and\ \bibinfo {author} {\bibfnamefont {P.}~\bibnamefont
  {{Slan{\'y}}}},\ }\href {\doibase 10.1103/PhysRevD.69.064001} {\bibfield
  {journal} {\bibinfo  {journal} {Phys. Rev. D}\ }\textbf {\bibinfo {volume}
  {69}},\ \bibinfo {eid} {064001} (\bibinfo {year} {2004})},\ \Eprint
  {http://arxiv.org/abs/gr-qc/0307049} {gr-qc/0307049} \BibitemShut {NoStop}%
\bibitem [{\citenamefont {{Stuchl{\'{\i}}k}}(2005)}]{Stuchlik05}%
  \BibitemOpen
  \bibfield  {author} {\bibinfo {author} {\bibfnamefont {Z.}~\bibnamefont
  {{Stuchl{\'{\i}}k}}},\ }\href {\doibase 10.1142/S0217732305016865} {\bibfield
   {journal} {\bibinfo  {journal} {Modern Physics Letters A}\ }\textbf
  {\bibinfo {volume} {20}},\ \bibinfo {pages} {561} (\bibinfo {year} {2005})},\
  \Eprint {http://arxiv.org/abs/0804.2266} {arXiv:0804.2266} \BibitemShut
  {NoStop}%
\bibitem [{\citenamefont {{Stuchl{\'{\i}}k}}\ and\ \citenamefont {{Kov{\'a}{\v
  r}}}(2008)}]{Stuchlik08}%
  \BibitemOpen
  \bibfield  {author} {\bibinfo {author} {\bibfnamefont {Z.}~\bibnamefont
  {{Stuchl{\'{\i}}k}}}\ and\ \bibinfo {author} {\bibfnamefont {J.}~\bibnamefont
  {{Kov{\'a}{\v r}}}},\ }\href {\doibase 10.1142/S021827180801373X} {\bibfield
  {journal} {\bibinfo  {journal} {International Journal of Modern Physics D}\
  }\textbf {\bibinfo {volume} {17}},\ \bibinfo {pages} {2089} (\bibinfo {year}
  {2008})},\ \Eprint {http://arxiv.org/abs/0803.3641} {arXiv:0803.3641 [gr-qc]}
  \BibitemShut {NoStop}%
\bibitem [{\citenamefont {{Stuchl{\'\i}k}}(2008)}]{Stuchlik2008arXiv}%
  \BibitemOpen
  \bibfield  {author} {\bibinfo {author} {\bibfnamefont {Z.}~\bibnamefont
  {{Stuchl{\'\i}k}}},\ }\href@noop {} {\bibfield  {journal} {\bibinfo
  {journal} {arXiv e-prints}\ ,\ \bibinfo {eid} {arXiv:0803.2530}} (\bibinfo
  {year} {2008})},\ \Eprint {http://arxiv.org/abs/0803.2530} {arXiv:0803.2530
  [gr-qc]} \BibitemShut {NoStop}%
\bibitem [{\citenamefont {{Stuchl{\'{\i}}k}}\ \emph {et~al.}(2009)\citenamefont
  {{Stuchl{\'{\i}}k}}, \citenamefont {{Slan{\'y}}},\ and\ \citenamefont
  {{Kov{\'a}{\v r}}}}]{Stuchlik09a}%
  \BibitemOpen
  \bibfield  {author} {\bibinfo {author} {\bibfnamefont {Z.}~\bibnamefont
  {{Stuchl{\'{\i}}k}}}, \bibinfo {author} {\bibfnamefont {P.}~\bibnamefont
  {{Slan{\'y}}}}, \ and\ \bibinfo {author} {\bibfnamefont {J.}~\bibnamefont
  {{Kov{\'a}{\v r}}}},\ }\href {\doibase 10.1088/0264-9381/26/21/215013}
  {\bibfield  {journal} {\bibinfo  {journal} {Classical and Quantum Gravity}\
  }\textbf {\bibinfo {volume} {26}},\ \bibinfo {eid} {215013} (\bibinfo {year}
  {2009})},\ \Eprint {http://arxiv.org/abs/0910.3184} {arXiv:0910.3184 [gr-qc]}
  \BibitemShut {NoStop}%
\bibitem [{\citenamefont {{Stuchl{\'{\i}}k}}\ \emph {et~al.}(2016)\citenamefont
  {{Stuchl{\'{\i}}k}}, \citenamefont {{Hled{\'{\i}}k}},\ and\ \citenamefont
  {{Novotn{\'y}}}}]{Stuchlik16a}%
  \BibitemOpen
  \bibfield  {author} {\bibinfo {author} {\bibfnamefont {Z.}~\bibnamefont
  {{Stuchl{\'{\i}}k}}}, \bibinfo {author} {\bibfnamefont {S.}~\bibnamefont
  {{Hled{\'{\i}}k}}}, \ and\ \bibinfo {author} {\bibfnamefont {J.}~\bibnamefont
  {{Novotn{\'y}}}},\ }\href {\doibase 10.1103/PhysRevD.94.103513} {\bibfield
  {journal} {\bibinfo  {journal} {Phys. Rev. D}\ }\textbf {\bibinfo {volume}
  {94}},\ \bibinfo {eid} {103513} (\bibinfo {year} {2016})},\ \Eprint
  {http://arxiv.org/abs/1611.05327} {arXiv:1611.05327 [gr-qc]} \BibitemShut
  {NoStop}%
\bibitem [{\citenamefont {{Charbul{\'a}k}}\ and\ \citenamefont
  {{Stuchl{\'\i}k}}(2017)}]{Charbulak17}%
  \BibitemOpen
  \bibfield  {author} {\bibinfo {author} {\bibfnamefont {D.}~\bibnamefont
  {{Charbul{\'a}k}}}\ and\ \bibinfo {author} {\bibfnamefont {Z.}~\bibnamefont
  {{Stuchl{\'\i}k}}},\ }\href {\doibase 10.1140/epjc/s10052-017-5401-9}
  {\bibfield  {journal} {\bibinfo  {journal} {European Physical Journal C}\
  }\textbf {\bibinfo {volume} {77}},\ \bibinfo {eid} {897} (\bibinfo {year}
  {2017})},\ \Eprint {http://arxiv.org/abs/1702.07850} {arXiv:1702.07850
  [gr-qc]} \BibitemShut {NoStop}%
\bibitem [{\citenamefont {{Stuchl{\'\i}k}}\ \emph {et~al.}(2018)\citenamefont
  {{Stuchl{\'\i}k}}, \citenamefont {{Charbul{\'a}k}},\ and\ \citenamefont
  {{Schee}}}]{Stuchlik18}%
  \BibitemOpen
  \bibfield  {author} {\bibinfo {author} {\bibfnamefont {Z.}~\bibnamefont
  {{Stuchl{\'\i}k}}}, \bibinfo {author} {\bibfnamefont {D.}~\bibnamefont
  {{Charbul{\'a}k}}}, \ and\ \bibinfo {author} {\bibfnamefont {J.}~\bibnamefont
  {{Schee}}},\ }\href {\doibase 10.1140/epjc/s10052-018-5578-6} {\bibfield
  {journal} {\bibinfo  {journal} {European Physical Journal C}\ }\textbf
  {\bibinfo {volume} {78}},\ \bibinfo {eid} {180} (\bibinfo {year} {2018})},\
  \Eprint {http://arxiv.org/abs/1811.00072} {arXiv:1811.00072 [gr-qc]}
  \BibitemShut {NoStop}%
\bibitem [{\citenamefont {{Bardeen}}\ \emph {et~al.}(1972)\citenamefont
  {{Bardeen}}, \citenamefont {{Press}},\ and\ \citenamefont
  {{Teukolsky}}}]{Bardeen72}%
  \BibitemOpen
  \bibfield  {author} {\bibinfo {author} {\bibfnamefont {J.~M.}\ \bibnamefont
  {{Bardeen}}}, \bibinfo {author} {\bibfnamefont {W.~H.}\ \bibnamefont
  {{Press}}}, \ and\ \bibinfo {author} {\bibfnamefont {S.~A.}\ \bibnamefont
  {{Teukolsky}}},\ }\href {\doibase 10.1086/151796} {\bibfield  {journal}
  {\bibinfo  {journal} {Astrophys. J.}\ }\textbf {\bibinfo {volume} {178}},\
  \bibinfo {pages} {347} (\bibinfo {year} {1972})}\BibitemShut {NoStop}%
\end{thebibliography}%
\end{document}